\documentstyle[12pt,piclatex,epsf]{article}
\newdimen\xposition \newdimen\yposition \newdimen\dyposition
\newdimen\crossbarlength \crossbarlength=5pt
\def\errbar #1 #2 #3 {%
 \xposition=\Xdistance{#1} \yposition=\Ydistance{#2}
 \dyposition=\Ydistance{#3}
 \setdimensionmode
 \dimen0 = \yposition \advance \dimen0 by -\dyposition
 \dimen2 = \yposition \advance \dimen2 by  \dyposition
 \putrule from {\xposition} {\dimen0} to {\xposition} {\dimen2}
 \dimen4 = \xposition \advance \dimen4 by -.5\crossbarlength
 \dimen6 = \xposition \advance \dimen6 by  .5\crossbarlength
 \putrule from {\dimen4} {\dimen0} to {\dimen6} {\dimen0}
 \putrule from {\dimen4} {\dimen2} to {\dimen6} {\dimen2}
 \setcoordinatemode}
\makeatletter
\@addtoreset{equation}{section}
\makeatother

\def\lb{\hfil\penalty-10000}

\begin{document}
\thispagestyle{empty}
\noindent \hspace{1cm} June 1993 \hfill HLRZ\,-\,93-39 \hspace{1cm}\\
\mbox{} \hfill BI-TP 93/21 \hspace{1cm}\\
\begin{center}
\vspace*{1.0cm}
{\large \bf Thermal Fluctuations of \\
\rule[-7pt]{0pt}{19pt}Chern-Simons Numbers \\
in the Lattice SU(2) Higgs Model} \\
\vspace*{1.0cm}
{\large F.\,Karsch$^{1,2}$, M.\,L.\,Laursen$^{1}$, \\
T.\,Neuhaus$^{2}$ and B.\,Plache$^{2}$} \\
\vspace*{1.0cm}
{\normalsize
$\mbox{}^1$ {HLRZ, c/o KFA J\"{u}lich, D-52425 J\"ulich, Germany}\\
$\mbox{}^2$ {Fak. f. Physik, Univ. Bielefeld, P.\,O.\,Box 100131,
D-33501 Bielefeld, Germany}}\\
\vspace*{2cm}
{\large \bf Abstract}
\end{center}
\setlength{\baselineskip}{1.3\baselineskip}

We study the temperature dependence of the Chern-Simons number fluctuations
in the $SU(2)$ Higgs Model on Euclidean lattices with spatial sizes up
to $20^3$.
Tempera\-tures well above the Higgs phase transition $T_H$ are achieved on
anisotropic lattices.
Numerical results are compared to perturbative results on finite
lattices as well as in continuum perturbation theory.
We find qualitative agreement with perturbative estimates and see at
high temperatures a tendency towards static configurations.
Up to temperatures $T \simeq 2T_H$ we find indication that tunneling
between
vacuums with different Chern-Simons numbers are still exponentially
suppressed.
\newpage
\setcounter{page}{1}
\section{Introduction}

The non-conservation of the baryon and lepton number in the electroweak
theory is well known \cite{Hooft}. It has been argued that the baryon
number violating processes can be strongly enhanced at high temperature.
Calculations of the corresponding rates are based on the
one hand on
semi-classical estimates for transitions between topologically distinct
vacuums of the electroweak theory \cite{ArnMcLe},
on the other hand they have been performed through Monte Carlo
simulations within the
framework of an effective, Hamiltonian model, which is derived from the
finite temperature Euclidean theory in the high temperature limit
\cite{Ambjorn}.
Both approaches rely on dimensional reduction, which is expected to
be valid at high temperature and should allow to treat the timelike
component of the $SU(2)$ gauge fields as static fields.
In the vicinity of the electroweak phase transition
this approach breaks down and a
non-perturbative
understanding of the structure of the topologically distinct vacuums and
their temperature dependence will be important.
We will study here thermal fluctuations of Chern-Simons number
distributions within the framework of the Euclidean formulation of the
$SU(2)$ Higgs model on the lattice. From an analysis of their
correlation in Euclidean time we will be able to test the validity of
the static approximation in the vicinity of the electroweak phase
transition. The non-perturbative results for this correlations as well
as moments of the Chern-Simons numbers will be compared with
high-temperature perturbation theory in the continuum and
on the lattice.

A baryon number violating process is a transition from a vacuum
field configuration to one in
another vacuum, which can not be reached through small gauge
transformations from the initial one.
In the $A_0=0$ gauge it is possible to assign a Chern-Simons number,
\begin{equation}
n_{cs} = -\frac{1}{8\pi^2} \int d^3\!x \ \epsilon_{ijk}
\ tr\left(A_i\left(\partial_jA_k+{\textstyle\frac{2}{3}}A_jA_k\right)
  \right)\,, \label{EQ-CS}
\end{equation}
to gauge field configurations at some fixed time.
Vacuums always have integer $n_{cs}$, and the change in baryon number
during the transition
between two of them is proportional to the difference of their
Chern-Simons numbers, the proportionality constant being the number of
quark and lepton families. The rate of transitions between different
vacuum sectors depends of the height of the potential barrier between
them. For temperatures below the electroweak phase transition, the rate for
baryon number violating processes can be estimated, making use of the existence
of classical field configurations, {\it sphalerons},
which interpolate between two vacuums and carry half-integer Chern-Simons
number. This approximation breaks down close to $T_c$. For temperatures much
larger than $T_c$ baryon number changing processes are expected to be frequent,
$\sim T^4$, however no accurate estimate exist. In particular in the vicinity
of $T_c$ the transition rates in the high temperature phase are unknown.
A non-perturbative study of the different topological
vacuums in the electroweak theory thus is needed.

In general the barrier height between vacuums with different Chern-Simons
numbers
can be determined from the difference of the Chern-Simons effective
potential $V(n_{cs})$ at $n_{cs}=0$ and $n_{cs}=1/2$, which is defined
as
\begin{displaymath}
P(n_{cs}) = e^{V(n_{cs})} = \int [D\!A][D\phi]\ e^{-S}\ \delta\left( n_{cs}\!+
\frac{1}{8\pi^2} \int\!d^3\!x \ \epsilon_{ijk}
\ tr\left(A_i(\partial_jA_k+ {\textstyle\frac{2}{3}}A_jA_k)\right)\right).
\label{pot}
\end{displaymath}
Here $P(n_{cs})$ is the probability distribution of the
Chern-Simons numbers.
The shape of the effective potential has been studied in the semi-classical
approach \cite{Akiba}.
Having a lattice prescription for the
evaluation \cite{Talks,Karsch} of $P(n_{cs})$ at hand,
one can extract the barrier height also nonperturbatively. The
temperature dependence of the barrier height can be used to determine the
temperature dependence of the transition rate, up to a normalization factor
\cite{Kripfganz}. Before one can aim to study this systematically for the
$SU(2)$ Higgs model on the lattice, it seems to us that a detailed study
of basic features of the Chern-Simons number distributions on the lattice
is needed. We started this investigation in a previous paper \cite{Karsch}
where we described in detail the lattice version of the Chern-Simons number and
studied their distributions on finite lattices. It is the purpose of the
present paper to investigate the temperature dependence of the Chern-Simons
number distributions. For small thermal fluctuations around vacuum
configurations
this has been studied in the continuum \cite{Shap}. We have repeated this
analytic $O(g^2)$ calculation on finite, anisotropic lattices and will compare
with Monte Carlo data for moments of the Chern-Simons distribution functions.
 From the calculated probability for the occurrence of configurations with
$n_{cs}=1/2$ we will obtain an
estimate for the temperature dependence of the tunnel rate between
different vacuums.

\pagebreak

This paper is organized as follows. In the next section we describe the
formulation
of finite temperature $SU(2)$ Higgs model on anisotropic lattices, which is
needed to study this model in the high temperature phase on a lattice with
finite
cut-off. In section 3 we discuss the perturbative calculation of thermal
fluctuations of $n_{cs}$ in the continuum and on the lattice. In section 4 we
present our numerical data and compare with perturbative results. We will also
study the correlation between Chern-Simons numbers extracted from different
time
slices of the (3+1)-dimensional lattice. This will give some hints on the
validity of the static approximation used in analytic calculations.
Finally we give our conclusions in section 5.

\section{Higgs Model on Anisotropic Lattices}

The thermodynamics of the $SU(2)$ Higgs model has been studied recently
in a detailed simulation at small bare gauge coupling,
$\beta = 4/g^2=8.0$, and
quartic self-coupling, $\lambda=0.0017235$, which have been chosen such
that the Higgs
boson mass is approximately equal to that of the W-boson \cite{Bunk}.
For these parameters a weak first order phase transition has been
found and the critical value of the hopping parameter,
$\kappa$, in the standard lattice action,
\begin{equation}
S = -\frac{\beta}{2} \sum_{n,\mu <\nu} tr(U_{n,\mu\nu})
-\kappa\sum_{n,\mu}tr(\phi_n^{\dag} U_{n,\mu} \phi_{n+\mu})
+\lambda\sum_n(\phi_n^{\dag}\phi_{n}-1)^2,
\label{action}
\end{equation}
has been determined to be $\kappa_c=0.12996$ on a lattice with temporal
extent $N_{\tau}=2$. We will use this set of couplings in our
simulations of the high temperature phase. In order to be able to go to
even higher temperatures than the critical temperature, $T_c$, on a
lattice of temporal extent $N_{\tau}=2$, we will perform simulations on
anisotropic lattices \cite{FK,Stam}. We thus replace Eq.\,(\ref{action}) by
\begin{eqnarray}
S= & {\displaystyle -\frac{1}{2}\beta_{\tau} \sum_{n,\nu=1,2,3} tr(U_{n,0\nu})
-\frac{1}{2}\beta_{\sigma} \sum_{n,1\le\mu<\nu} tr(U_{n,\mu\nu}) }
\nonumber \\
 & {\displaystyle -\kappa_{\tau}\sum_{n}tr(\phi_n^{\dag} U_{n,0}\phi_{n+0})
 -\kappa_{\sigma} \sum_{n,\mu=1,2,3} tr(\phi_n^{\dag} U_{n,\mu} \phi_{n+0}) }
\nonumber \\
  &{\displaystyle +\lambda\sum_n(\phi_n^{\dag}\phi_{n}-1)^2.}
\label{aniso}
\end{eqnarray}
Choosing different values for the spacelike $(\sigma)$ and timelike
$(\tau)$ couplings allows to vary the corresponding lattice spacings,
$a_\sigma$ and $a_\tau$, independently. In order to insure Lorentz invariance
in
the continuum limit, the couplings can, however, not be changed
independently from
each other. A common parametrization is
\begin{eqnarray}
& & \beta_\sigma \!=\!\beta /\gamma_g \qquad , \qquad
\beta_\tau\!=\! \beta\gamma_g \qquad
\nonumber \\
& & \kappa_\sigma\!=\!\kappa/\gamma_h \qquad , \qquad
\kappa_\tau\!=\!\kappa\gamma_h\ . \qquad
\label{anisocoup}
\end{eqnarray}
In a leading order weak coupling expansion one finds
\begin{equation}
\gamma_g=\gamma_h= \xi\!=\!a_\sigma/a_\tau ~~~.
\end{equation}
To this order the anisotropy couplings $\gamma_{g,h}$ thus directly give the
anisotropy in the lattice spacings.
Corrections to these relations are $O(\beta^{-1})$ and have been found to
be smaller than $10\%$ for the $SU(2)$ Higgs model \cite{Aniso}.
On a lattice with anisotropy $\xi$ the temperature, measured in units of
the spatial lattice cut-off, $1/a_{\sigma}$, thus is given by
\begin{equation}
a_{\sigma}T = \frac{\xi}{N_{\tau}}~~~.
\label{temper}
\end{equation}
We have performed simulations with different $N_{\tau}$ and $\xi$,
but constant ratio, $\xi/N_{\tau}=a_{\sigma}T$, in order to check that
within our numerical accuracy results
indeed depend only on this ratio and not on $N_\tau$ and $\xi$
separately.

\section{Perturbation Theory}

The calculation of Chern-Simons numbers on the lattice is not
straightforward. A characteristic feature of the continuum expression,
Eq.\,(\ref{EQ-CS}), is that it changes under gauge
transformations only by integers. The non-integer part of $n_{cs}$ thus
is gauge invariant.
A naive discretization of the continuum expression,
Eq.\,(\ref{EQ-CS}), would not have this property. It is, however, possible to
give a geometric definition for $n_{cs}$ on the lattice, which is
closely related to the geometric definitions of the topological charge
\cite{Luesch,Seiberg} and which still retains the transformation
properties of the continuum expression. We have examined this
formulation in detail in previous publications \cite{Talks,Karsch}. In
particular we have shown there that the geometric definition indeed
leads to a periodic probability distribution for the Chern-Simons
numbers and that an accurate calculation of the gauge
invariant non-integer part of $n_{cs}$ is possible even in the case of
rather rough field configuration, which occur for small values of
$\beta$.
Due to the periodicity of the Chern-Simons potential it is sufficient to
consider then
only the gauge invariant non-integer part of $n_{cs}$,
which we normalize such that $n_{cs} \in [-1/2,1/2]$.

The thermal fluctuations of Chern-Simons numbers, Eq.\,(1.1), have been
evaluated in continuum
perturbation theory to leading order in the gauge coupling $g^2$ \cite{Shap},
\begin{equation}
\langle n_{cs}^2 \rangle =\left( \frac{g^2}{32\pi^2}\right)^2
\int d^3x \int d^3y
\langle \varepsilon_{ijk} F_{ij}^a(x) A_k^a(x)
\varepsilon_{lmn} F_{lm}^b(y) A_n^b(y) \rangle~~,
\label{Ncscontinuum}
\end{equation}
where $\langle ... \rangle$ denotes thermal expectation values calculated with
the quadratic part of the Euclidean finite temperature Lagrangian.
The Chern-Simons number distribution and its width is an extensive
quantity, while different Chern-Simons vacuums are only separated by
integers. In the large volume limit even small local quantum or thermal
fluctuations of the gauge fields can add up to large Chern-Simons
numbers and will lead to a flat distribution. Also a perturbative
evaluation of the width of the thermal distribution thus only makes
sense in a finite volume. A leading order perturbative evaluation of
Eq.\,(\ref{Ncscontinuum}) in a volume of size $V\!=\!L^3$ yields
\begin{equation}
\langle n_{cs}^2 \rangle_T=6\left( \frac{g^2}{32\pi^2}\right)^2
\sum_{n_1,n_2,n_3=0,\pm1,...}\bigl(n_B+n_B^2\bigr)~~,
\label{ncsf}
\end{equation}
where the Bose-distribution function, $n_B$, in a finite volume at
temperature $T$ is given by
$n_B=1/(\exp(\sqrt{n_1^2+n_2^2+n_3^2}/LT-1)$. In Eq.\,(3.2) the contribution
from zero temperature, vacuum fluctuations has been subtracted.
In the limit of large values of $LT$ the sums in Eq.\,(\ref{ncsf}) can be
converted into a three-dimensional integral and an explicit evaluation
of this yields\footnote{
The large volume limit has previously been evaluated in Ref.\,\cite{Shap}.
The result given there is too large by a factor 2 \cite{Shaprev}.
Besides this our result differs from Ref.\,\cite{Shap} in the way the vacuum
subtraction is performed.
This results in a factor $\zeta(2)$ rather than $\zeta(2)-\zeta(3)$ as
found in Ref.\,\cite{Shap}.}
\begin{equation}
\langle n_{cs}^2 \rangle_T
= \frac{24VT^3}{\pi^2} \zeta(2)
\left( \frac{g^2}{32\pi^2}\right)^2
= 4VT^3\left( \frac{g^2}{32\pi^2}\right)^2 \ .
\label{ncs}
\end{equation}
For small values of $LT$ this asymptotic formula overestimates the size
of thermal fluctuations.
We find, however, that Eq.\,(\ref{ncs}) approximates the exact result,
Eq.\,(\ref{ncsf}), better than within 10\% for $LT > 16$.

\begin{figure}
\beginpicture

\setcoordinatesystem units <1pt,1pt>
\setcoordinatesystem point at 0 0
\setplotarea x from -10 to 360, y from 0 to 40

\put { \beginpicture

\setcoordinatesystem units <12pt,0.005pt>
\setcoordinatesystem point at 0 0
\setplotarea x from 0 to 22, y from 0 to 33000

\axis left ticks in withvalues {$1\!\cdot\!10^4$} {$2\!\cdot\!10^4$}
{$3\!\cdot\!10^4$} / at 10000 20000 30000 / /
\put {${\displaystyle \frac{\langle n_{cs}^2\rangle}{(g^2/32\pi^2)^2}}$}
[r] at -0.5 34000

\axis bottom ticks in numbered at 5 10 15 / /
\put {$LT$} at 19 -2500

\setquadratic \linethickness=.4pt

\setsolid
\plot  0 0 1 4 2 32 3 108 4 256 5 500 6 864 7 1372 8 2048 9 2916 10 4000
      11 5324 12 6912 13 8788 14 10976 15 13500 16 16384 17 19652 18 23328
      19 27436 20 32000 /
\setdashes
\plot  0 0 1 0.3145 2 12.321 3 61.230 4 171.30 5 366.53 6 670.92 7 1108.5
       8 1703.2 9 2479.1 10 3460.1 11 4670.3 12 6133.7 13 7874.2 14 9915.9
      15 12283 16 14999 17 18088 18 21574 19 25482 20 29834 /
\setdots
\plot  2 8.9731 3 68.966 4 204.97 5 436.92 6 774.06 7 1218.6 8 1769.1 9 2422.9
      10 3177.1 11 4028.9 12 4976.3 13 6017.3 14 7150.5 15 8374.9 16 9689.4
      17 11093 18 12586 19 14168 20 15837 /
\setdashpattern <2pt,4pt>
\plot  2 9.1121 3 76.948 4 236.57 5 506.42 6 888.54 7 1378.8 8 1972.0 9 2663.9
      10 3451.2 11 4331.6 12 5303.4 13 6365.2 14 7516.3 15 8755.9 16 10084
      17 11499 18 13002 19 14592 20 16269 /
\endpicture
} [lb] at 0 0

\endpicture

\caption{\sl
The dashed curve shows the continuum perturbative result, Eq.\,(3.2),
while the solid line is the large volume approximation, Eq.\,(3.3).
The lower dotted curve has been evaluated on $2\times 8^3$ lattices
using Eq.\,(3.7). It shows strong deviations for large values of
$LT=N_\sigma\xi/N_\tau$;
the lattice result is seen to change only slightly in the limit
$\xi\!\rightarrow\!\infty\,,N_\tau\!\rightarrow\!\infty$,
$\xi/N_\tau\!=\!$ const.\ (short dashes). }
\end{figure}

We have evaluated the corresponding expression on a finite Euclidean lattice
of size $N_{\tau} \times N_{\sigma}^3$ with anisotropy $\xi$. Using the
standard perturbative expressions for the gauge field propagators on the
lattice \cite{Heller}, we find
\begin{equation}
\langle n_{cs}^2 \rangle_\xi (N_\sigma, N_\tau) = \frac{6\xi^2}{N_\tau^2}
\left( \frac{g^2}{32\pi^2} \right)^2 \sum_{\vec p,p_0,q_0}'
\frac{ S^2_{\vec p} }{ (S^2_{\vec p} + \xi^2 \sin^2(p_0/2) )
(S^2_{\vec p} + \xi^2 \sin^2(q_0/2) ) }~~,
\label{latsum}
\end{equation}
where $ S^2_{\vec p} =\sin^2(p_1/2) + \sin^2(p_2/2) + \sin^2(p_3/2)$
and where the momenta take on the values
\begin{eqnarray}
p_i & = & 2\pi n_\sigma / N_\sigma \quad , \quad i=1,2,3\nonumber \\
p_0, q_0 & = & 2\pi n_\tau / N_\tau
\quad , \quad n_{\sigma,\tau} = 0,\,1,\,\dots,N_{\sigma,\tau}-1~~.
\end{eqnarray}
The prime on the sum appearing in Eq.\,(\ref{latsum}) indicates that the zero
mode does not contribute. The thermal part of the Chern-Simons number
fluctuations is obtained by subtracting the zero temperature contribution,
defined through Eq.\,(\ref{latsum}) in the limit $N_\tau \rightarrow \infty$
with $N_\sigma$ and $\xi$ fixed,
\begin{equation}
\langle n_{cs}^2 \rangle_\xi (N_\sigma,\infty)
=6\,\xi^2 \left( \frac{g^2}{32\pi^2} \right)^2
\sum_{\vec p}'{1 \over \xi^2 + S^2_{\vec p}} ~~.
\label{latsum0}
\end{equation}
The lattice analog to Eq.\,(\ref{ncs}) thus is obtained as
\begin{equation}
\langle n_{cs}^2 \rangle_T=
\langle n_{cs}^2 \rangle_\xi (N_\sigma, N_\tau)-
\langle n_{cs}^2 \rangle_\xi (N_\sigma, \infty)~~.
\label{ncslat}
\end{equation}

A comparison of the perturbative results on the lattice
and in the continuum is shown in Fig.\,1.
The agreement of the perturbative result on a lattice of spatial
extent $N_{\sigma}=8$ with the corresponding result from continuum
perturbation theory ($LT=N_\sigma \xi / N_\tau$),
is already remarkably good for $N_\tau =2$ and $\xi \le 2.5$, ie. $LT
\le 10$. We note, however, that
changing the temperature by varying $\xi$ and keeping $N_\tau$ fixed is
possible only in a limited parameter range; it will lead
to a wrong temperature dependence ($\xi$-dependence) for
large values of $\xi$ as can be seen from Eq.\,(\ref{latsum}).
We have to take the continuum limit by taking $N_\sigma ,N_\tau$ and
$\xi$ to infinity while $\xi / N_\tau$ is kept fixed.

We also give here the analog expression for correlations between
Chern-Simons numbers on different (Euclidean) time slices. The basic assumption
in analytic continuum calculations, as well as real time simulations on the
lattice \cite{Ambjorn,Shap} is that at high temperature the relevant
field
configurations become static. This should be reflected in an increasing
correlation between Chern-Simons numbers on neighboring time slices.
The perturbative calculation yields,
\begin{equation}
\langle n_{cs}(0) n_{cs}(t) \rangle =\frac{6\,\xi^2}{N_\tau^2}
\left( \frac{g^2}{32\pi^2} \right)^2 \sum_{\vec p,p_0,q_0}'
\frac{ S^2_{\vec p} \cos(q_0t)} { (S^2_{\vec p} + \xi^2 \sin^2(p_0/2) )
(S^2_{\vec p} + \xi^2 \sin^2(q_0/2) ) }~~.
\end{equation}

In the following we will use these perturbative expressions for comparison
with our Monte Carlo data. This will allow to judge, in how far the thermal
fluctuations in the Chern-Simons numbers show perturbative behaviour in the
temperature and coupling regime studied by us.

\section{Numerical Results}

All our simulations have been performed with the set of couplings
$\beta\!=\!4/g^2\!=\!8$, $\kappa\!=\!0.12996$ and $\lambda\!=\!0.0017235$
\cite{Bunk},
and various values of the anisotropy $\xi \in [1,2.5]$. The finite temperature
simulations have generally been performed on lattices of size $N_{\tau} \times
N_{\sigma}^3$ with $N_{\tau}=2$ and $N_{\sigma}$ ranging from 4 to 20.
Some simulations
have been performed on lattices with larger values of $N_{\tau}$ in order to
check that within our statistical accuracy our finite temperature
results only depend on the ratio $a_{\sigma}T= \xi/N_{\tau}$.
For the zero temperature subtractions we use symmetric lattices
of size $N_{\sigma}^4$ with the same value of $\xi$.
We note that we will calculate only the gauge invariant non-integer part
of $n_{cs}$, which is normalized such that $n_{cs} \in [-1/2,1/2]$.
In the case of a flat distribution of Chern-Simons numbers, which will
be reached in the limit
of large volumes or temperatures, one thus will find the limiting value
\begin{equation}
\langle n_{cs}^2 \rangle_\xi (N_\sigma, N_\tau)
\frac{}{N_\sigma , \xi/N_\tau\, \rightarrow \infty\ \ } \!\!\!\!\!\! >
\ \frac{1}{12}~~~.
\label{limit}
\end{equation}
This constrains the range of useful lattice sizes for our simulations.
However, it also indicates that the perturbative calculations are only valid
for small values of $LT$.
Taking the perturbative results as a first guidance,
we find that for $g^2 = 0.5$, as we will use it in our simulations%
\footnote{The
small value of $g^2$, used in this study, has an additional advantage over
our earlier investigation \cite{Karsch}:
Due to the smoothness of the gauge fields the evaluation of Chern-Simons
numbers by numerical integration is much simplified.
This also improves the speed of our numerical algorithm, which to a
large extent vectorizes and on a CRAY-YMP or NEC SX-3 typically
yields one Chern-Simons number in 40 sec.\ on a $2\times 10^3$ lattice.},
the asymptotic value, $\langle n_{cs}^2 \rangle=1/12$ will be reached for
$LT= N_\sigma \xi/N_\tau \simeq 20$. As will become
clear from the following discussion we, indeed, find that the limit given by
Eq.\,(\ref{limit}) is reached for $LT \simeq 10$, thus for $N_\tau =2$ we can
work on rather large spatial lattices up to size $N_\sigma = 20$

Some distributions of Chern-Simons numbers on isotropic
lattices $(\xi\!=\!1)$ are shown in the
first two rows of Fig.\,2.
We note that with increasing spatial lattice size the distributions get
significantly broader.
However, as expected we also find that in this regime of couplings large values
of $n_{cs}$, in the vicinity of $n_{cs}=1/2$, are strongly suppressed at zero
temperature and even close to $T_c$ (second row in Fig.\,2), although
the somewhat broader distributions found in this latter case suggest
an increasing tunneling probability compared to that at zero
temperature.

\begin{figure}
\def\PR{\putrule from }
\beginpicture

\setcoordinatesystem units <1pt,1pt>
\setcoordinatesystem point at 0 0
\setplotarea x from -10 to 360, y from 0 to 40

\put { \beginpicture
\setcoordinatesystem units <120pt,6pt> point at 0 0
\setplotarea x from -0.75 to 0.5, y from 0 to 25
\axis left shiftedto x=-0.5 invisible ticks in numbered at 5 10 15 20 / /
\put {$\rho_{cs}$} [r] at -0.55 23.2
\axis right  invisible ticks in at 5 10 15 20 / /
\axis bottom invisible ticks in at -0.25 0 0.25 / /
\PR -0.5 25 to -0.5  0  \PR  0.5  0 to  0.5 25    \PR  0.5 25 to -0.5 25
\setlinear  \linethickness=.4pt

\PR -0.09 0 to -0.08750 0.00000
\PR -0.0875 0.00000 to -0.0875 0.02381
\PR -0.0875 0.02381 to -0.0825 0.02381
\PR -0.0825 0.02381 to -0.0825 0.0
\PR -0.0825 0.00000 to -0.0775 0.0
\PR -0.0775 0.00000 to -0.0775 0.02381
\PR -0.0775 0.02381 to -0.0725 0.02381
\PR -0.0725 0.02381 to -0.0725 0.04762
\PR -0.0725 0.04762 to -0.0675 0.04762
\PR -0.0675 0.04762 to -0.0675 0.04762
\PR -0.0675 0.04762 to -0.0625 0.04762
\PR -0.0625 0.04762 to -0.0625 0.19048
\PR -0.0625 0.19048 to -0.0575 0.19048
\PR -0.0575 0.19048 to -0.0575 0.33333
\PR -0.0575 0.33333 to -0.0525 0.33333
\PR -0.0525 0.33333 to -0.0525 0.54762
\PR -0.0525 0.54762 to -0.0475 0.54762
\PR -0.0475 0.54762 to -0.0475 0.76190
\PR -0.0475 0.76190 to -0.0425 0.76190
\PR -0.0425 0.76190 to -0.0425 1.76190
\PR -0.0425 1.76190 to -0.0375 1.76190
\PR -0.0375 1.76190 to -0.0375 3.21429
\PR -0.0375 3.21429 to -0.0325 3.21429
\PR -0.0325 3.21429 to -0.0325 5.97619
\PR -0.0325 5.97619 to -0.0275 5.97619
\PR -0.0275 5.97619 to -0.0275 8.92857
\PR -0.0275 8.92857 to -0.0225 8.92857
\PR -0.0225 8.92857 to -0.0225 11.76190
\PR -0.0225 11.76190 to -0.0175 11.76190
\PR -0.0175 11.76190 to -0.0175 14.0
\PR -0.0175 14.00000 to -0.0125 14.0
\PR -0.0125 14.00000 to -0.0125 19.04762
\PR -0.0125 19.04762 to -0.0075 19.04762
\PR -0.0075 19.04762 to -0.0075 21.57143
\PR -0.0075 21.57143 to -0.0025 21.57143
\PR -0.0025 21.57143 to -0.0025 23.11905
\PR -0.0025 23.11905 to 0.0025 23.11905
\PR 0.0025 23.11905 to 0.0025 20.88095
\PR 0.0025 20.88095 to 0.0075 20.88095
\PR 0.0075 20.88095 to 0.0075 18.42857
\PR 0.0075 18.42857 to 0.0125 18.42857
\PR 0.0125 18.42857 to 0.0125 15.64286
\PR 0.0125 15.64286 to 0.0175 15.64286
\PR 0.0175 15.64286 to 0.0175 11.83333
\PR 0.0175 11.83333 to 0.0225 11.83333
\PR 0.0225 11.83333 to 0.0225 8.45238
\PR 0.0225 8.45238 to 0.0275 8.45238
\PR 0.0275 8.45238 to 0.0275 5.73810
\PR 0.0275 5.73810 to 0.0325 5.73810
\PR 0.0325 5.73810 to 0.0325 3.47619
\PR 0.0325 3.47619 to 0.0375 3.47619
\PR 0.0375 3.47619 to 0.0375 1.92857
\PR 0.0375 1.92857 to 0.0425 1.92857
\PR 0.0425 1.92857 to 0.0425 1.21429
\PR 0.0425 1.21429 to 0.0475 1.21429
\PR 0.0475 1.21429 to 0.0475 0.59524
\PR 0.0475 0.59524 to 0.0525 0.59524
\PR 0.0525 0.59524 to 0.0525 0.26190
\PR 0.0525 0.26190 to 0.0575 0.26190
\PR 0.0575 0.26190 to 0.0575 0.14286
\PR 0.0575 0.14286 to 0.0625 0.14286
\PR 0.0625 0.14286 to 0.0625 0.04762
\PR 0.0625 0.04762 to 0.0675 0.04762
\PR 0.0675 0.04762 to 0.0675 0.0
\PR 0.0675 0.00000 to 0.0700 0.0

\put {$4^4, \xi=1$} at 0.22 23
\endpicture
} [lb] at -30 306

\put { \beginpicture
\setcoordinatesystem units <120pt,6pt> point at 0 0
\setplotarea x from -0.5 to 0.5, y from 0 to 25
\axis left   invisible ticks in at 5 10 15 20 / /
\axis right  invisible ticks in at 5 10 15 20 / /
\axis bottom invisible ticks in at -0.25 0 0.25 / /
\PR  0.5  0 to  0.5 25 \PR  0.5 25 to -0.5 25
\setlinear  \linethickness=.4pt

\PR -0.14 0.0 to -0.13 0.0
\PR -0.13 0.0 to -0.13 0.02033
\PR -0.13 0.02033 to -0.11 0.02033
\PR -0.11 0.02033 to -0.11 0.18293
\PR -0.11 0.18293 to -0.09 0.18293
\PR -0.09 0.18293 to -0.09 0.89431
\PR -0.09 0.89431 to -0.07 0.89431
\PR -0.07 0.89431 to -0.07 2.66260
\PR -0.07 2.66260 to -0.05 2.66260
\PR -0.05 2.66260 to -0.05 5.04065
\PR -0.05 5.04065 to -0.03 5.04065
\PR -0.03 5.04065 to -0.03 10.16260
\PR -0.03 10.1626 to -0.01 10.16260
\PR -0.01 10.1626 to -0.01 11.78862
\PR -0.01 11.78862 to 0.01 11.78862
\PR 0.01 11.78862 to 0.01 9.34959
\PR 0.01 9.34959 to 0.03 9.34959
\PR 0.03 9.34959 to 0.03 5.83333
\PR 0.03 5.83333 to 0.05 5.83333
\PR 0.05 5.83333 to 0.05 2.78455
\PR 0.05 2.78455 to 0.07 2.78455
\PR 0.07 2.78455 to 0.07 0.89431
\PR 0.07 0.89431 to 0.09 0.89431
\PR 0.09 0.89431 to 0.09 0.30488
\PR 0.09 0.30488 to 0.11 0.30488
\PR 0.11 0.30488 to 0.11 0.06098
\PR 0.11 0.06098 to 0.13 0.06098
\PR 0.13 0.06098 to 0.13 0.02033
\PR 0.13 0.02033 to 0.15 0.02033
\PR 0.15 0.02033 to 0.15 0.0
\PR 0.15 0.00000 to 0.16 0.0

\put {$6^4,\xi=1$} at 0 23
\endpicture
} [lb] at 120 306

\put { \beginpicture
\setcoordinatesystem units <120pt,6pt> point at 0 0
\setplotarea x from -0.5 to 0.75, y from 0 to 25
\axis left  invisible ticks in at 5 10 15 20 / /
\axis right shiftedto x=0.5 invisible ticks in numbered at 5 10 15 20 / /
\put {$\rho_{cs}$} [l] at  0.55 23.2
\axis bottom invisible ticks in at -0.25 0 0.25 / /
\PR  0.5  0 to  0.5 25 \PR  0.5 25 to -0.5 25
\setlinear  \linethickness=.4pt

\PR -0.19333 0 to -0.18667 0.00000
\PR -0.18667 0.00000 to -0.18667 0.06466
\PR -0.18667 0.06466 to -0.17333 0.06466
\PR -0.17333 0.06466 to -0.17333 0.16164
\PR -0.17333 0.16164 to -0.16000 0.16164
\PR -0.16000 0.16164 to -0.16000 0.12931
\PR -0.16000 0.12931 to -0.14667 0.12931
\PR -0.14667 0.12931 to -0.14667 0.22629
\PR -0.14667 0.22629 to -0.13333 0.22629
\PR -0.13333 0.22629 to -0.13333 0.32328
\PR -0.13333 0.32328 to -0.12000 0.32328
\PR -0.12000 0.32328 to -0.12000 0.77586
\PR -0.12000 0.77586 to -0.10667 0.77586
\PR -0.10667 0.77586 to -0.10667 1.09914
\PR -0.10667 1.09914 to -0.09333 1.09914
\PR -0.09333 1.09914 to -0.09333 2.03664
\PR -0.09333 2.03664 to -0.08000 2.03664
\PR -0.08000 2.03664 to -0.08000 2.81250
\PR -0.08000 2.81250 to -0.06667 2.81250
\PR -0.06667 2.81250 to -0.06667 3.62069
\PR -0.06667 3.62069 to -0.05333 3.62069
\PR -0.05333 3.62069 to -0.05333 5.04310
\PR -0.05333 5.04310 to -0.04000 5.04310
\PR -0.04000 5.04310 to -0.04000 6.30388
\PR -0.04000 6.30388 to -0.02667 6.30388
\PR -0.02667 6.30388 to -0.02667 7.07974
\PR -0.02667 7.07974 to -0.01333 7.07974
\PR -0.01333 7.07974 to -0.01333 7.69397
\PR -0.01333 7.69397 to 0.00000 7.69397
\PR 0.00000 7.69397 to 0.00000 6.62716
\PR 0.00000 6.62716 to 0.01333 6.62716
\PR 0.01333 6.62716 to 0.01333 7.30603
\PR 0.01333 7.30603 to 0.02667 7.30603
\PR 0.02667 7.30603 to 0.02667 5.49569
\PR 0.02667 5.49569 to 0.04000 5.49569
\PR 0.04000 5.49569 to 0.04000 5.49569
\PR 0.04000 5.49569 to 0.05333 5.49569
\PR 0.05333 5.49569 to 0.05333 4.39655
\PR 0.05333 4.39655 to 0.06667 4.39655
\PR 0.06667 4.39655 to 0.06667 3.26509
\PR 0.06667 3.26509 to 0.08000 3.26509
\PR 0.08000 3.26509 to 0.08000 2.23060
\PR 0.08000 2.23060 to 0.09333 2.23060
\PR 0.09333 2.23060 to 0.09333 1.29310
\PR 0.09333 1.29310 to 0.10667 1.29310
\PR 0.10667 1.29310 to 0.10667 0.84052
\PR 0.10667 0.84052 to 0.12000 0.84052
\PR 0.12000 0.84052 to 0.12000 0.29095
\PR 0.12000 0.29095 to 0.13333 0.29095
\PR 0.13333 0.29095 to 0.13333 0.22629
\PR 0.13333 0.22629 to 0.14667 0.22629
\PR 0.14667 0.22629 to 0.14667 0.12931
\PR 0.14667 0.12931 to 0.16000 0.12931
\PR 0.16000 0.12931 to 0.16000 0.03233
\PR 0.16000 0.03233 to 0.17333 0.03233
\PR 0.17333 0.03233 to 0.17333 0.00000
\PR 0.17333 0.00000 to 0.18000 0.00000

\put {$8^4,\xi=1$} at 0 23
\endpicture
} [lb] at 240 306


\put { \beginpicture
\setcoordinatesystem units <120pt,6pt> point at 0 0
\setplotarea x from -0.75 to 0.5, y from 0 to 25
\axis left shiftedto x=-0.5 invisible ticks in numbered at 5 10 15 20 / /
\put {$\rho_{cs}$} [r] at -0.55 23.2
\axis right invisible ticks in at 5 10 15 20 / /
\axis bottom invisible ticks in at -0.25 0 0.25 / /
\PR -0.5 25 to -0.5  0 \PR  0.5  0 to  0.5 25 \PR  0.5 25 to -0.5 25
\setlinear  \linethickness=.4pt

\PR -0.0875 0 to -0.08687 0.00000
\PR -0.08687 0.00000 to -0.08687 0.08130
\PR -0.08687 0.08130 to -0.08563 0.08130
\PR -0.08563 0.08130 to -0.08563 0.00000
\PR -0.08563 0.00000 to -0.08438 0.00000
\PR -0.08438 0.00000 to -0.08438 0.00000
\PR -0.08438 0.00000 to -0.08313 0.00000
\PR -0.08313 0.00000 to -0.08313 0.00000
\PR -0.08313 0.00000 to -0.08188 0.00000
\PR -0.08188 0.00000 to -0.08188 0.00000
\PR -0.08188 0.00000 to -0.08063 0.00000
\PR -0.08063 0.00000 to -0.08063 0.00000
\PR -0.08063 0.00000 to -0.07938 0.00000
\PR -0.07938 0.00000 to -0.07938 0.00000
\PR -0.07938 0.00000 to -0.07812 0.00000
\PR -0.07812 0.00000 to -0.07812 0.00000
\PR -0.07812 0.00000 to -0.07687 0.00000
\PR -0.07687 0.00000 to -0.07687 0.08130
\PR -0.07687 0.08130 to -0.07562 0.08130
\PR -0.07562 0.08130 to -0.07562 0.00000
\PR -0.07562 0.00000 to -0.07437 0.00000
\PR -0.07437 0.00000 to -0.07437 0.00000
\PR -0.07437 0.00000 to -0.07312 0.00000
\PR -0.07312 0.00000 to -0.07312 0.00000
\PR -0.07312 0.00000 to -0.07187 0.00000
\PR -0.07187 0.00000 to -0.07187 0.00000
\PR -0.07187 0.00000 to -0.07062 0.00000
\PR -0.07062 0.00000 to -0.07062 0.16260
\PR -0.07062 0.16260 to -0.06938 0.16260
\PR -0.06938 0.16260 to -0.06938 0.24390
\PR -0.06938 0.24390 to -0.06813 0.24390
\PR -0.06813 0.24390 to -0.06813 0.32520
\PR -0.06813 0.32520 to -0.06688 0.32520
\PR -0.06688 0.32520 to -0.06688 0.32520
\PR -0.06688 0.32520 to -0.06563 0.32520
\PR -0.06563 0.32520 to -0.06563 0.24390
\PR -0.06563 0.24390 to -0.06438 0.24390
\PR -0.06438 0.24390 to -0.06438 0.16260
\PR -0.06438 0.16260 to -0.06313 0.16260
\PR -0.06313 0.16260 to -0.06313 0.48780
\PR -0.06313 0.48780 to -0.06187 0.48780
\PR -0.06187 0.48780 to -0.06187 0.73171
\PR -0.06187 0.73171 to -0.06062 0.73171
\PR -0.06062 0.73171 to -0.06062 0.40650
\PR -0.06062 0.40650 to -0.05937 0.40650
\PR -0.05937 0.40650 to -0.05937 0.65041
\PR -0.05937 0.65041 to -0.05812 0.65041
\PR -0.05812 0.65041 to -0.05812 0.40650
\PR -0.05812 0.40650 to -0.05688 0.40650
\PR -0.05688 0.40650 to -0.05688 0.32520
\PR -0.05688 0.32520 to -0.05563 0.32520
\PR -0.05563 0.32520 to -0.05563 0.40650
\PR -0.05563 0.40650 to -0.05437 0.40650
\PR -0.05437 0.40650 to -0.05437 0.73171
\PR -0.05437 0.73171 to -0.05312 0.73171
\PR -0.05312 0.73171 to -0.05312 0.81301
\PR -0.05312 0.81301 to -0.05187 0.81301
\PR -0.05187 0.81301 to -0.05187 0.97561
\PR -0.05187 0.97561 to -0.05063 0.97561
\PR -0.05063 0.97561 to -0.05063 1.13821
\PR -0.05063 1.13821 to -0.04938 1.13821
\PR -0.04938 1.13821 to -0.04938 1.21951
\PR -0.04938 1.21951 to -0.04813 1.21951
\PR -0.04813 1.21951 to -0.04813 1.21951
\PR -0.04813 1.21951 to -0.04688 1.21951
\PR -0.04688 1.21951 to -0.04688 1.62602
\PR -0.04688 1.62602 to -0.04562 1.62602
\PR -0.04562 1.62602 to -0.04562 2.43902
\PR -0.04562 2.43902 to -0.04437 2.43902
\PR -0.04437 2.43902 to -0.04437 2.27642
\PR -0.04437 2.27642 to -0.04312 2.27642
\PR -0.04312 2.27642 to -0.04312 2.03252
\PR -0.04312 2.03252 to -0.04188 2.03252
\PR -0.04188 2.03252 to -0.04188 3.08943
\PR -0.04188 3.08943 to -0.04063 3.08943
\PR -0.04063 3.08943 to -0.04063 2.60163
\PR -0.04063 2.60163 to -0.03938 2.60163
\PR -0.03938 2.60163 to -0.03938 3.41463
\PR -0.03938 3.41463 to -0.03812 3.41463
\PR -0.03812 3.41463 to -0.03812 2.43902
\PR -0.03812 2.43902 to -0.03687 2.43902
\PR -0.03687 2.43902 to -0.03687 4.95935
\PR -0.03687 4.95935 to -0.03563 4.95935
\PR -0.03563 4.95935 to -0.03563 5.52846
\PR -0.03563 5.52846 to -0.03438 5.52846
\PR -0.03438 5.52846 to -0.03438 5.85366
\PR -0.03438 5.85366 to -0.03313 5.85366
\PR -0.03313 5.85366 to -0.03313 5.77236
\PR -0.03313 5.77236 to -0.03188 5.77236
\PR -0.03188 5.77236 to -0.03188 5.69106
\PR -0.03188 5.69106 to -0.03062 5.69106
\PR -0.03062 5.69106 to -0.03062 8.61789
\PR -0.03062 8.61789 to -0.02937 8.61789
\PR -0.02937 8.61789 to -0.02937 7.07317
\PR -0.02937 7.07317 to -0.02813 7.07317
\PR -0.02813 7.07317 to -0.02813 7.15447
\PR -0.02813 7.15447 to -0.02687 7.15447
\PR -0.02687 7.15447 to -0.02687 9.10569
\PR -0.02687 9.10569 to -0.02563 9.10569
\PR -0.02563 9.10569 to -0.02563 9.10569
\PR -0.02563 9.10569 to -0.02438 9.10569
\PR -0.02438 9.10569 to -0.02438 9.26829
\PR -0.02438 9.26829 to -0.02312 9.26829
\PR -0.02312 9.26829 to -0.02312 9.26829
\PR -0.02312 9.26829 to -0.02187 9.26829
\PR -0.02187 9.26829 to -0.02187 10.81301
\PR -0.02187 10.81301 to -0.02063 10.81301
\PR -0.02063 10.81301 to -0.02063 13.08943
\PR -0.02063 13.08943 to -0.01937 13.08943
\PR -0.01937 13.08943 to -0.01937 12.60163
\PR -0.01937 12.60163 to -0.01813 12.60163
\PR -0.01813 12.60163 to -0.01813 12.84553
\PR -0.01813 12.84553 to -0.01688 12.84553
\PR -0.01688 12.84553 to -0.01688 14.39024
\PR -0.01688 14.39024 to -0.01562 14.39024
\PR -0.01562 14.39024 to -0.01562 13.33333
\PR -0.01562 13.33333 to -0.01437 13.33333
\PR -0.01437 13.33333 to -0.01437 12.27642
\PR -0.01437 12.27642 to -0.01313 12.27642
\PR -0.01313 12.27642 to -0.01313 15.20325
\PR -0.01313 15.20325 to -0.01188 15.20325
\PR -0.01188 15.20325 to -0.01188 17.31707
\PR -0.01188 17.31707 to -0.01062 17.31707
\PR -0.01062 17.31707 to -0.01062 17.64228
\PR -0.01062 17.64228 to -0.00938 17.64228
\PR -0.00938 17.64228 to -0.00938 18.86179
\PR -0.00938 18.86179 to -0.00813 18.86179
\PR -0.00813 18.86179 to -0.00813 17.88618
\PR -0.00813 17.88618 to -0.00688 17.88618
\PR -0.00688 17.88618 to -0.00688 17.80488
\PR -0.00688 17.80488 to -0.00562 17.80488
\PR -0.00562 17.80488 to -0.00562 16.66667
\PR -0.00562 16.66667 to -0.00438 16.66667
\PR -0.00438 16.66667 to -0.00438 20.48780
\PR -0.00438 20.48780 to -0.00313 20.48780
\PR -0.00313 20.48780 to -0.00313 19.91870
\PR -0.00313 19.91870 to -0.00187 19.91870
\PR -0.00187 19.91870 to -0.00187 19.67480
\PR -0.00187 19.67480 to -0.00063 19.67480
\PR -0.00063 19.67480 to -0.00063 19.91870
\PR -0.00063 19.91870 to 0.00063 19.91870
\PR 0.00063 19.91870 to 0.00063 20.16260
\PR 0.00063 20.16260 to 0.00187 20.16260
\PR 0.00187 20.16260 to 0.00187 20.97561
\PR 0.00187 20.97561 to 0.00313 20.97561
\PR 0.00313 20.97561 to 0.00313 19.83740
\PR 0.00313 19.83740 to 0.00438 19.83740
\PR 0.00438 19.83740 to 0.00438 19.02439
\PR 0.00438 19.02439 to 0.00562 19.02439
\PR 0.00562 19.02439 to 0.00562 17.47967
\PR 0.00562 17.47967 to 0.00688 17.47967
\PR 0.00688 17.47967 to 0.00688 18.45528
\PR 0.00688 18.45528 to 0.00813 18.45528
\PR 0.00813 18.45528 to 0.00813 16.74797
\PR 0.00813 16.74797 to 0.00938 16.74797
\PR 0.00938 16.74797 to 0.00938 17.72358
\PR 0.00938 17.72358 to 0.01062 17.72358
\PR 0.01062 17.72358 to 0.01062 16.66667
\PR 0.01062 16.66667 to 0.01188 16.66667
\PR 0.01188 16.66667 to 0.01188 16.42276
\PR 0.01188 16.42276 to 0.01313 16.42276
\PR 0.01313 16.42276 to 0.01313 12.92683
\PR 0.01313 12.92683 to 0.01437 12.92683
\PR 0.01437 12.92683 to 0.01437 15.04065
\PR 0.01437 15.04065 to 0.01562 15.04065
\PR 0.01562 15.04065 to 0.01562 13.25203
\PR 0.01562 13.25203 to 0.01688 13.25203
\PR 0.01688 13.25203 to 0.01688 11.86992
\PR 0.01688 11.86992 to 0.01813 11.86992
\PR 0.01813 11.86992 to 0.01813 12.68293
\PR 0.01813 12.68293 to 0.01937 12.68293
\PR 0.01937 12.68293 to 0.01937 12.35772
\PR 0.01937 12.35772 to 0.02063 12.35772
\PR 0.02063 12.35772 to 0.02063 11.05691
\PR 0.02063 11.05691 to 0.02187 11.05691
\PR 0.02187 11.05691 to 0.02187 11.54472
\PR 0.02187 11.54472 to 0.02312 11.54472
\PR 0.02312 11.54472 to 0.02312 10.73171
\PR 0.02312 10.73171 to 0.02438 10.73171
\PR 0.02438 10.73171 to 0.02438 10.56911
\PR 0.02438 10.56911 to 0.02563 10.56911
\PR 0.02563 10.56911 to 0.02563 8.21138
\PR 0.02563 8.21138 to 0.02687 8.21138
\PR 0.02687 8.21138 to 0.02687 7.31707
\PR 0.02687 7.31707 to 0.02813 7.31707
\PR 0.02813 7.31707 to 0.02813 7.07317
\PR 0.02813 7.07317 to 0.02937 7.07317
\PR 0.02937 7.07317 to 0.02937 7.72358
\PR 0.02937 7.72358 to 0.03062 7.72358
\PR 0.03062 7.72358 to 0.03062 5.36585
\PR 0.03062 5.36585 to 0.03188 5.36585
\PR 0.03188 5.36585 to 0.03188 5.52846
\PR 0.03188 5.52846 to 0.03313 5.52846
\PR 0.03313 5.52846 to 0.03313 5.20325
\PR 0.03313 5.20325 to 0.03438 5.20325
\PR 0.03438 5.20325 to 0.03438 3.65854
\PR 0.03438 3.65854 to 0.03563 3.65854
\PR 0.03563 3.65854 to 0.03563 4.63415
\PR 0.03563 4.63415 to 0.03687 4.63415
\PR 0.03687 4.63415 to 0.03687 4.06504
\PR 0.03687 4.06504 to 0.03812 4.06504
\PR 0.03812 4.06504 to 0.03812 2.60163
\PR 0.03812 2.60163 to 0.03938 2.60163
\PR 0.03938 2.60163 to 0.03938 2.35772
\PR 0.03938 2.35772 to 0.04063 2.35772
\PR 0.04063 2.35772 to 0.04063 2.35772
\PR 0.04063 2.35772 to 0.04188 2.35772
\PR 0.04188 2.35772 to 0.04188 2.35772
\PR 0.04188 2.35772 to 0.04312 2.35772
\PR 0.04312 2.35772 to 0.04312 2.76423
\PR 0.04312 2.76423 to 0.04437 2.76423
\PR 0.04437 2.76423 to 0.04437 2.03252
\PR 0.04437 2.03252 to 0.04562 2.03252
\PR 0.04562 2.03252 to 0.04562 1.30081
\PR 0.04562 1.30081 to 0.04688 1.30081
\PR 0.04688 1.30081 to 0.04688 1.54472
\PR 0.04688 1.54472 to 0.04813 1.54472
\PR 0.04813 1.54472 to 0.04813 1.62602
\PR 0.04813 1.62602 to 0.04938 1.62602
\PR 0.04938 1.62602 to 0.04938 0.65041
\PR 0.04938 0.65041 to 0.05063 0.65041
\PR 0.05063 0.65041 to 0.05063 0.73171
\PR 0.05063 0.73171 to 0.05187 0.73171
\PR 0.05187 0.73171 to 0.05187 1.21951
\PR 0.05187 1.21951 to 0.05312 1.21951
\PR 0.05312 1.21951 to 0.05312 0.40650
\PR 0.05312 0.40650 to 0.05437 0.40650
\PR 0.05437 0.40650 to 0.05437 0.65041
\PR 0.05437 0.65041 to 0.05563 0.65041
\PR 0.05563 0.65041 to 0.05563 0.40650
\PR 0.05563 0.40650 to 0.05688 0.40650
\PR 0.05688 0.40650 to 0.05688 0.40650
\PR 0.05688 0.40650 to 0.05812 0.40650
\PR 0.05812 0.40650 to 0.05812 0.40650
\PR 0.05812 0.40650 to 0.05937 0.40650
\PR 0.05937 0.40650 to 0.05937 0.65041
\PR 0.05937 0.65041 to 0.06062 0.65041
\PR 0.06062 0.65041 to 0.06062 0.16260
\PR 0.06062 0.16260 to 0.06187 0.16260
\PR 0.06187 0.16260 to 0.06187 0.16260
\PR 0.06187 0.16260 to 0.06313 0.16260
\PR 0.06313 0.16260 to 0.06313 0.40650
\PR 0.06313 0.40650 to 0.06438 0.40650
\PR 0.06438 0.40650 to 0.06438 0.32520
\PR 0.06438 0.32520 to 0.06563 0.32520
\PR 0.06563 0.32520 to 0.06563 0.16260
\PR 0.06563 0.16260 to 0.06688 0.16260
\PR 0.06688 0.16260 to 0.06688 0.16260
\PR 0.06688 0.16260 to 0.06813 0.16260
\PR 0.06813 0.16260 to 0.06813 0.00000
\PR 0.06813 0.00000 to 0.06938 0.00000
\PR 0.06938 0.00000 to 0.06938 0.16260
\PR 0.06938 0.16260 to 0.07062 0.16260
\PR 0.07062 0.16260 to 0.07062 0.16260
\PR 0.07062 0.16260 to 0.07187 0.16260
\PR 0.07187 0.16260 to 0.07187 0.00000
\PR 0.07187 0.00000 to 0.07312 0.00000
\PR 0.07312 0.00000 to 0.07312 0.00000
\PR 0.07312 0.00000 to 0.07437 0.00000
\PR 0.07437 0.00000 to 0.07437 0.16260
\PR 0.07437 0.16260 to 0.07562 0.16260
\PR 0.07562 0.16260 to 0.07562 0.08130
\PR 0.07562 0.08130 to 0.07687 0.08130
\PR 0.07687 0.08130 to 0.07687 0.00000
\PR 0.07687 0.00000 to 0.07812 0.00000
\PR 0.07812 0.00000 to 0.07812 0.08130
\PR 0.07812 0.08130 to 0.07938 0.08130
\PR 0.07938 0.08130 to 0.07938 0.00000
\PR 0.07938 0.00000 to 0.08000 0.00000

\put {$2\times 4^3,\xi=1$} at 0 23
\endpicture
} [lb] at -30 156

\put { \beginpicture
\setcoordinatesystem units <120pt,6pt> point at 0 0
\setplotarea x from -0.5 to 0.5, y from 0 to 25
\axis left  invisible ticks in at 5 10 15 20 / /
\axis right invisible ticks in at 5 10 15 20 / /
\axis bottom invisible ticks in at -0.25 0 0.25 / /
\PR  0.5  0 to  0.5 25 \PR  0.5 25 to -0.5 25
\setlinear  \linethickness=.4pt

\PR -0.165 0 to -0.16375 0.00000
\PR -0.16375 0.00000 to -0.16375 0.05208
\PR -0.16375 0.05208 to -0.16125 0.05208
\PR -0.16125 0.05208 to -0.16125 0.00000
\PR -0.14625 0.00000 to -0.14625 0.05208
\PR -0.14625 0.05208 to -0.14375 0.05208
\PR -0.14375 0.05208 to -0.14375 0.05208
\PR -0.14375 0.05208 to -0.14125 0.05208
\PR -0.14125 0.05208 to -0.14125 0.00000
\PR -0.14125 0.00000 to -0.13875 0.00000
\PR -0.13875 0.00000 to -0.13875 0.05208
\PR -0.13875 0.05208 to -0.13625 0.05208
\PR -0.13625 0.05208 to -0.13625 0.00000
\PR -0.13375 0.00000 to -0.13375 0.10417
\PR -0.13375 0.10417 to -0.13125 0.10417
\PR -0.13125 0.10417 to -0.13125 0.00000
\PR -0.13125 0.00000 to -0.12875 0.00000
\PR -0.12875 0.00000 to -0.12875 0.10417
\PR -0.12875 0.10417 to -0.12625 0.10417
\PR -0.12625 0.10417 to -0.12625 0.05208
\PR -0.12625 0.05208 to -0.12375 0.05208
\PR -0.12375 0.05208 to -0.12375 0.15625
\PR -0.12375 0.15625 to -0.12125 0.15625
\PR -0.12125 0.15625 to -0.12125 0.05208
\PR -0.12125 0.05208 to -0.11875 0.05208
\PR -0.11875 0.05208 to -0.11875 0.10417
\PR -0.11875 0.10417 to -0.11625 0.10417
\PR -0.11625 0.10417 to -0.11625 0.36458
\PR -0.11625 0.36458 to -0.11375 0.36458
\PR -0.11375 0.36458 to -0.11375 0.10417
\PR -0.11375 0.10417 to -0.11125 0.10417
\PR -0.11125 0.10417 to -0.11125 0.31250
\PR -0.11125 0.31250 to -0.10875 0.31250
\PR -0.10875 0.31250 to -0.10875 0.15625
\PR -0.10875 0.15625 to -0.10625 0.15625
\PR -0.10625 0.15625 to -0.10625 0.31250
\PR -0.10625 0.31250 to -0.10375 0.31250
\PR -0.10375 0.31250 to -0.10375 0.57292
\PR -0.10375 0.57292 to -0.10125 0.57292
\PR -0.10125 0.57292 to -0.10125 0.88542
\PR -0.10125 0.88542 to -0.09875 0.88542
\PR -0.09875 0.88542 to -0.09875 0.72917
\PR -0.09875 0.72917 to -0.09625 0.72917
\PR -0.09625 0.72917 to -0.09625 0.98958
\PR -0.09625 0.98958 to -0.09375 0.98958
\PR -0.09375 0.98958 to -0.09375 0.98958
\PR -0.09375 0.98958 to -0.09125 0.98958
\PR -0.09125 0.98958 to -0.09125 1.19792
\PR -0.09125 1.19792 to -0.08875 1.19792
\PR -0.08875 1.19792 to -0.08875 1.25000
\PR -0.08875 1.25000 to -0.08625 1.25000
\PR -0.08625 1.25000 to -0.08625 0.67708
\PR -0.08625 0.67708 to -0.08375 0.67708
\PR -0.08375 0.67708 to -0.08375 1.77083
\PR -0.08375 1.77083 to -0.08125 1.77083
\PR -0.08125 1.77083 to -0.08125 1.40625
\PR -0.08125 1.40625 to -0.07875 1.40625
\PR -0.07875 1.40625 to -0.07875 1.45833
\PR -0.07875 1.45833 to -0.07625 1.45833
\PR -0.07625 1.45833 to -0.07625 2.03125
\PR -0.07625 2.03125 to -0.07375 2.03125
\PR -0.07375 2.03125 to -0.07375 1.82292
\PR -0.07375 1.82292 to -0.07125 1.82292
\PR -0.07125 1.82292 to -0.07125 2.18750
\PR -0.07125 2.18750 to -0.06875 2.18750
\PR -0.06875 2.18750 to -0.06875 2.70833
\PR -0.06875 2.70833 to -0.06625 2.70833
\PR -0.06625 2.70833 to -0.06625 2.34375
\PR -0.06625 2.34375 to -0.06375 2.34375
\PR -0.06375 2.34375 to -0.06375 3.22917
\PR -0.06375 3.22917 to -0.06125 3.22917
\PR -0.06125 3.22917 to -0.06125 3.43750
\PR -0.06125 3.43750 to -0.05875 3.43750
\PR -0.05875 3.43750 to -0.05875 3.95833
\PR -0.05875 3.95833 to -0.05625 3.95833
\PR -0.05625 3.95833 to -0.05625 3.85417
\PR -0.05625 3.85417 to -0.05375 3.85417
\PR -0.05375 3.85417 to -0.05375 5.20833
\PR -0.05375 5.20833 to -0.05125 5.20833
\PR -0.05125 5.20833 to -0.05125 5.10417
\PR -0.05125 5.10417 to -0.04875 5.10417
\PR -0.04875 5.10417 to -0.04875 5.41667
\PR -0.04875 5.41667 to -0.04625 5.41667
\PR -0.04625 5.41667 to -0.04625 5.15625
\PR -0.04625 5.15625 to -0.04375 5.15625
\PR -0.04375 5.15625 to -0.04375 4.94792
\PR -0.04375 4.94792 to -0.04125 4.94792
\PR -0.04125 4.94792 to -0.04125 6.25000
\PR -0.04125 6.25000 to -0.03875 6.25000
\PR -0.03875 6.25000 to -0.03875 5.93750
\PR -0.03875 5.93750 to -0.03625 5.93750
\PR -0.03625 5.93750 to -0.03625 6.19792
\PR -0.03625 6.19792 to -0.03375 6.19792
\PR -0.03375 6.19792 to -0.03375 7.55208
\PR -0.03375 7.55208 to -0.03125 7.55208
\PR -0.03125 7.55208 to -0.03125 6.97917
\PR -0.03125 6.97917 to -0.02875 6.97917
\PR -0.02875 6.97917 to -0.02875 7.55208
\PR -0.02875 7.55208 to -0.02625 7.55208
\PR -0.02625 7.55208 to -0.02625 6.30208
\PR -0.02625 6.30208 to -0.02375 6.30208
\PR -0.02375 6.30208 to -0.02375 9.27083
\PR -0.02375 9.27083 to -0.02125 9.27083
\PR -0.02125 9.27083 to -0.02125 9.47917
\PR -0.02125 9.47917 to -0.01875 9.47917
\PR -0.01875 9.47917 to -0.01875 8.85417
\PR -0.01875 8.85417 to -0.01625 8.85417
\PR -0.01625 8.85417 to -0.01625 8.22917
\PR -0.01625 8.22917 to -0.01375 8.22917
\PR -0.01375 8.22917 to -0.01375 9.84375
\PR -0.01375 9.84375 to -0.01125 9.84375
\PR -0.01125 9.84375 to -0.01125 8.95833
\PR -0.01125 8.95833 to -0.00875 8.95833
\PR -0.00875 8.95833 to -0.00875 10.41667
\PR -0.00875 10.41667 to -0.00625 10.41667
\PR -0.00625 10.41667 to -0.00625 10.41667
\PR -0.00625 10.41667 to -0.00375 10.41667
\PR -0.00375 10.41667 to -0.00375 9.21875
\PR -0.00375 9.21875 to -0.00125 9.21875
\PR -0.00125 9.21875 to -0.00125 11.09375
\PR -0.00125 11.09375 to 0.00125 11.09375
\PR 0.00125 11.09375 to 0.00125 8.95833
\PR 0.00125 8.95833 to 0.00375 8.95833
\PR 0.00375 8.95833 to 0.00375 9.73958
\PR 0.00375 9.73958 to 0.00625 9.73958
\PR 0.00625 9.73958 to 0.00625 9.32292
\PR 0.00625 9.32292 to 0.00875 9.32292
\PR 0.00875 9.32292 to 0.00875 9.68750
\PR 0.00875 9.68750 to 0.01125 9.68750
\PR 0.01125 9.68750 to 0.01125 9.27083
\PR 0.01125 9.27083 to 0.01375 9.27083
\PR 0.01375 9.27083 to 0.01375 9.42708
\PR 0.01375 9.42708 to 0.01625 9.42708
\PR 0.01625 9.42708 to 0.01625 7.96875
\PR 0.01625 7.96875 to 0.01875 7.96875
\PR 0.01875 7.96875 to 0.01875 8.59375
\PR 0.01875 8.59375 to 0.02125 8.59375
\PR 0.02125 8.59375 to 0.02125 8.22917
\PR 0.02125 8.22917 to 0.02375 8.22917
\PR 0.02375 8.22917 to 0.02375 8.33333
\PR 0.02375 8.33333 to 0.02625 8.33333
\PR 0.02625 8.33333 to 0.02625 7.34375
\PR 0.02625 7.34375 to 0.02875 7.34375
\PR 0.02875 7.34375 to 0.02875 7.44792
\PR 0.02875 7.44792 to 0.03125 7.44792
\PR 0.03125 7.44792 to 0.03125 6.97917
\PR 0.03125 6.97917 to 0.03375 6.97917
\PR 0.03375 6.97917 to 0.03375 6.45833
\PR 0.03375 6.45833 to 0.03625 6.45833
\PR 0.03625 6.45833 to 0.03625 6.71875
\PR 0.03625 6.71875 to 0.03875 6.71875
\PR 0.03875 6.71875 to 0.03875 5.72917
\PR 0.03875 5.72917 to 0.04125 5.72917
\PR 0.04125 5.72917 to 0.04125 5.10417
\PR 0.04125 5.10417 to 0.04375 5.10417
\PR 0.04375 5.10417 to 0.04375 5.26042
\PR 0.04375 5.26042 to 0.04625 5.26042
\PR 0.04625 5.26042 to 0.04625 4.42708
\PR 0.04625 4.42708 to 0.04875 4.42708
\PR 0.04875 4.42708 to 0.04875 4.11458
\PR 0.04875 4.11458 to 0.05125 4.11458
\PR 0.05125 4.11458 to 0.05125 4.01042
\PR 0.05125 4.01042 to 0.05375 4.01042
\PR 0.05375 4.01042 to 0.05375 3.59375
\PR 0.05375 3.59375 to 0.05625 3.59375
\PR 0.05625 3.59375 to 0.05625 3.22917
\PR 0.05625 3.22917 to 0.05875 3.22917
\PR 0.05875 3.22917 to 0.05875 3.33333
\PR 0.05875 3.33333 to 0.06125 3.33333
\PR 0.06125 3.33333 to 0.06125 2.55208
\PR 0.06125 2.55208 to 0.06375 2.55208
\PR 0.06375 2.55208 to 0.06375 3.38542
\PR 0.06375 3.38542 to 0.06625 3.38542
\PR 0.06625 3.38542 to 0.06625 2.29167
\PR 0.06625 2.29167 to 0.06875 2.29167
\PR 0.06875 2.29167 to 0.06875 2.55208
\PR 0.06875 2.55208 to 0.07125 2.55208
\PR 0.07125 2.55208 to 0.07125 2.08333
\PR 0.07125 2.08333 to 0.07375 2.08333
\PR 0.07375 2.08333 to 0.07375 2.13542
\PR 0.07375 2.13542 to 0.07625 2.13542
\PR 0.07625 2.13542 to 0.07625 1.71875
\PR 0.07625 1.71875 to 0.07875 1.71875
\PR 0.07875 1.71875 to 0.07875 1.30208
\PR 0.07875 1.30208 to 0.08125 1.30208
\PR 0.08125 1.30208 to 0.08125 1.04167
\PR 0.08125 1.04167 to 0.08375 1.04167
\PR 0.08375 1.04167 to 0.08375 1.09375
\PR 0.08375 1.09375 to 0.08625 1.09375
\PR 0.08625 1.09375 to 0.08625 0.88542
\PR 0.08625 0.88542 to 0.08875 0.88542
\PR 0.08875 0.88542 to 0.08875 1.30208
\PR 0.08875 1.30208 to 0.09125 1.30208
\PR 0.09125 1.30208 to 0.09125 1.04167
\PR 0.09125 1.04167 to 0.09375 1.04167
\PR 0.09375 1.04167 to 0.09375 0.78125
\PR 0.09375 0.78125 to 0.09625 0.78125
\PR 0.09625 0.78125 to 0.09625 0.46875
\PR 0.09625 0.46875 to 0.09875 0.46875
\PR 0.09875 0.46875 to 0.09875 0.57292
\PR 0.09875 0.57292 to 0.10125 0.57292
\PR 0.10125 0.57292 to 0.10125 0.57292
\PR 0.10125 0.57292 to 0.10375 0.57292
\PR 0.10375 0.57292 to 0.10375 0.36458
\PR 0.10375 0.36458 to 0.10625 0.36458
\PR 0.10625 0.36458 to 0.10625 0.31250
\PR 0.10625 0.31250 to 0.10875 0.31250
\PR 0.10875 0.31250 to 0.10875 0.20833
\PR 0.10875 0.20833 to 0.11125 0.20833
\PR 0.11125 0.20833 to 0.11125 0.41667
\PR 0.11125 0.41667 to 0.11375 0.41667
\PR 0.11375 0.41667 to 0.11375 0.15625
\PR 0.11375 0.15625 to 0.11625 0.15625
\PR 0.11625 0.15625 to 0.11625 0.20833
\PR 0.11625 0.20833 to 0.11875 0.20833
\PR 0.11875 0.20833 to 0.11875 0.15625
\PR 0.11875 0.15625 to 0.12125 0.15625
\PR 0.12125 0.15625 to 0.12125 0.20833
\PR 0.12125 0.20833 to 0.12375 0.20833
\PR 0.12375 0.20833 to 0.12375 0.20833
\PR 0.12375 0.20833 to 0.12625 0.20833
\PR 0.12625 0.20833 to 0.12625 0.20833
\PR 0.12625 0.20833 to 0.12875 0.20833
\PR 0.12875 0.20833 to 0.12875 0.10417
\PR 0.12875 0.10417 to 0.13125 0.10417
\PR 0.13125 0.10417 to 0.13125 0.00000
\PR 0.13125 0.00000 to 0.13375 0.00000
\PR 0.13375 0.00000 to 0.13375 0.00000
\PR 0.13375 0.00000 to 0.13625 0.00000
\PR 0.13625 0.00000 to 0.13625 0.05208
\PR 0.13625 0.05208 to 0.13875 0.05208
\PR 0.13875 0.05208 to 0.13875 0.20833
\PR 0.13875 0.20833 to 0.14125 0.20833
\PR 0.14125 0.20833 to 0.14125 0.05208
\PR 0.14125 0.05208 to 0.14375 0.05208
\PR 0.14375 0.05208 to 0.14375 0.05208
\PR 0.14375 0.05208 to 0.14625 0.05208
\PR 0.14625 0.05208 to 0.14625 0.00000
\PR 0.14625 0.00000 to 0.14875 0.00000
\PR 0.14875 0.00000 to 0.14875 0.00000
\PR 0.14875 0.00000 to 0.15125 0.00000
\PR 0.15125 0.00000 to 0.15125 0.10417
\PR 0.15125 0.10417 to 0.15375 0.10417
\PR 0.15375 0.10417 to 0.15375 0.00000
\PR 0.15375 0.00000 to 0.15500 0.00000

\put {$2\times 6^3, \xi=1$} at 0 23
\endpicture
} [lb] at 120 156

\put { \beginpicture
\setcoordinatesystem units <120pt,6pt> point at 0 0
\setplotarea x from -0.5 to 0.75, y from 0 to 25
\axis left  invisible ticks in at 5 10 15 20 / /
\axis right shiftedto x=0.5 invisible ticks in numbered at 5 10 15 20 / /
\put {$\rho_{cs}$} [l] at  0.55 23.2
\axis bottom invisible ticks in at -0.25 0 0.25 / /
\PR  0.5  0 to  0.5 25 \PR  0.5 25 to -0.5 25
\setlinear  \linethickness=.4pt

\PR -0.33333 0.00000 to -0.33333 0.00822
\PR -0.33333 0.00822 to -0.32000 0.00822
\PR -0.32000 0.00822 to -0.32000 0.00000
\PR -0.25333 0.00000 to -0.25333 0.00822
\PR -0.25333 0.00822 to -0.24000 0.00822
\PR -0.24000 0.00822 to -0.24000 0.00822
\PR -0.24000 0.00822 to -0.22667 0.00822
\PR -0.22667 0.00822 to -0.22667 0.02467
\PR -0.22667 0.02467 to -0.21333 0.02467
\PR -0.21333 0.02467 to -0.21333 0.04112
\PR -0.21333 0.04112 to -0.20000 0.04112
\PR -0.20000 0.04112 to -0.20000 0.05757
\PR -0.20000 0.05757 to -0.18667 0.05757
\PR -0.18667 0.05757 to -0.18667 0.10691
\PR -0.18667 0.10691 to -0.17333 0.10691
\PR -0.17333 0.10691 to -0.17333 0.21382
\PR -0.17333 0.21382 to -0.16000 0.21382
\PR -0.16000 0.21382 to -0.16000 0.43586
\PR -0.16000 0.43586 to -0.14667 0.43586
\PR -0.14667 0.43586 to -0.14667 0.69079
\PR -0.14667 0.69079 to -0.13333 0.69079
\PR -0.13333 0.69079 to -0.13333 0.95395
\PR -0.13333 0.95395 to -0.12000 0.95395
\PR -0.12000 0.95395 to -0.12000 1.31579
\PR -0.12000 1.31579 to -0.10667 1.31579
\PR -0.10667 1.31579 to -0.10667 2.00658
\PR -0.10667 2.00658 to -0.09333 2.00658
\PR -0.09333 2.00658 to -0.09333 2.45888
\PR -0.09333 2.45888 to -0.08000 2.45888
\PR -0.08000 2.45888 to -0.08000 3.33059
\PR -0.08000 3.33059 to -0.06667 3.33059
\PR -0.06667 3.33059 to -0.06667 4.12829
\PR -0.06667 4.12829 to -0.05333 4.12829
\PR -0.05333 4.12829 to -0.05333 4.84375
\PR -0.05333 4.84375 to -0.04000 4.84375
\PR -0.04000 4.84375 to -0.04000 5.31250
\PR -0.04000 5.31250 to -0.02667 5.31250
\PR -0.02667 5.31250 to -0.02667 5.69079
\PR -0.02667 5.69079 to -0.01333 5.69079
\PR -0.01333 5.69079 to -0.01333 5.74836
\PR -0.01333 5.74836 to 0.00000 5.74836
\PR 0.00000 5.74836 to 0.00000 6.11020
\PR 0.00000 6.11020 to 0.01333 6.11020
\PR 0.01333 6.11020 to 0.01333 6.06086
\PR 0.01333 6.06086 to 0.02667 6.06086
\PR 0.02667 6.06086 to 0.02667 5.33717
\PR 0.02667 5.33717 to 0.04000 5.33717
\PR 0.04000 5.33717 to 0.04000 4.62993
\PR 0.04000 4.62993 to 0.05333 4.62993
\PR 0.05333 4.62993 to 0.05333 3.96382
\PR 0.05333 3.96382 to 0.06667 3.96382
\PR 0.06667 3.96382 to 0.06667 3.13322
\PR 0.06667 3.13322 to 0.08000 3.13322
\PR 0.08000 3.13322 to 0.08000 2.62336
\PR 0.08000 2.62336 to 0.09333 2.62336
\PR 0.09333 2.62336 to 0.09333 2.12171
\PR 0.09333 2.12171 to 0.10667 2.12171
\PR 0.10667 2.12171 to 0.10667 1.20888
\PR 0.10667 1.20888 to 0.12000 1.20888
\PR 0.12000 1.20888 to 0.12000 0.91283
\PR 0.12000 0.91283 to 0.13333 0.91283
\PR 0.13333 0.91283 to 0.13333 0.61678
\PR 0.13333 0.61678 to 0.14667 0.61678
\PR 0.14667 0.61678 to 0.14667 0.41118
\PR 0.14667 0.41118 to 0.16000 0.41118
\PR 0.16000 0.41118 to 0.16000 0.23849
\PR 0.16000 0.23849 to 0.17333 0.23849
\PR 0.17333 0.23849 to 0.17333 0.10691
\PR 0.17333 0.10691 to 0.18667 0.10691
\PR 0.18667 0.10691 to 0.18667 0.08224
\PR 0.18667 0.08224 to 0.20000 0.08224
\PR 0.20000 0.08224 to 0.20000 0.03289
\PR 0.20000 0.03289 to 0.21333 0.03289
\PR 0.21333 0.03289 to 0.21333 0.00822
\PR 0.21333 0.00822 to 0.22667 0.00822
\PR 0.22667 0.00822 to 0.22667 0.01645
\PR 0.22667 0.01645 to 0.24000 0.01645
\PR 0.24000 0.01645 to 0.24000 0.00000
  \put {$2\times 8^3,\xi=1$} at 0 23
  \endpicture
  } [lb] at 240 156


\put { \beginpicture
\setcoordinatesystem units <120pt,24pt> point at 0 0
\setplotarea x from -0.75 to 0.5, y from -1 to 6.5
\axis left shiftedto x=-0.5 invisible ticks in numbered  at 2 4 /
    short unlabeled in at 1 3 5 / /
\put {$\rho_{cs}$} [r] at -0.55 5.85
\axis right invisible ticks in at 2 4 / short in unlabeled at 1 3 5 / /
\axis bottom shiftedto y=0 invisible ticks in numbered at -0.25 0 0.25 / /
\PR -0.5 6.5 to -0.5 0          \PR -0.5   0 to  0.5 0
\PR  0.5   0 to  0.5 6.5        \PR  0.5 6.5 to -0.5 6.5
\setlinear  \linethickness=.4pt

\PR -0.47 0.00000 to -0.47 0.00504
\PR -0.47 0.00504 to -0.45 0.00504
\PR -0.45 0.00504 to -0.45 0.00000
\PR -0.41 0.00000 to -0.41 0.00504
\PR -0.41 0.00504 to -0.39 0.00504
\PR -0.39 0.00504 to -0.39 0.01008
\PR -0.39 0.01008 to -0.37 0.01008
\PR -0.37 0.01008 to -0.37 0.01008
\PR -0.37 0.01008 to -0.35 0.01008
\PR -0.35 0.01008 to -0.35 0.02015
\PR -0.35 0.02015 to -0.33 0.02015
\PR -0.33 0.02015 to -0.33 0.00504
\PR -0.33 0.00504 to -0.31 0.00504
\PR -0.31 0.00504 to -0.31 0.03023
\PR -0.31 0.03023 to -0.29 0.03023
\PR -0.29 0.03023 to -0.29 0.06046
\PR -0.29 0.06046 to -0.27 0.06046
\PR -0.27 0.06046 to -0.27 0.09573
\PR -0.27 0.09573 to -0.25 0.09573
\PR -0.25 0.09573 to -0.25 0.24184
\PR -0.25 0.24184 to -0.23 0.24184
\PR -0.23 0.24184 to -0.23 0.34764
\PR -0.23 0.34764 to -0.21 0.34764
\PR -0.21 0.34764 to -0.21 0.52902
\PR -0.21 0.52902 to -0.19 0.52902
\PR -0.19 0.52902 to -0.19 0.70536
\PR -0.19 0.70536 to -0.17 0.70536
\PR -0.17 0.70536 to -0.17 1.12858
\PR -0.17 1.12858 to -0.15 1.12858
\PR -0.15 1.12858 to -0.15 1.43087
\PR -0.15 1.43087 to -0.13 1.43087
\PR -0.13 1.43087 to -0.13 2.01028
\PR -0.13 2.01028 to -0.11 2.01028
\PR -0.11 2.01028 to -0.11 2.50907
\PR -0.11 2.50907 to -0.09 2.50907
\PR -0.09 2.50907 to -0.09 2.81137
\PR -0.09 2.81137 to -0.07 2.81137
\PR -0.07 2.81137 to -0.07 3.46634
\PR -0.07 3.46634 to -0.05 3.46634
\PR -0.05 3.46634 to -0.05 3.67291
\PR -0.05 3.67291 to -0.03 3.67291
\PR -0.03 3.67291 to -0.03 4.04575
\PR -0.03 4.04575 to -0.01 4.04575
\PR -0.01 4.04575 to -0.01 4.11628
\PR -0.01 4.11628 to 0.01 4.11628
\PR 0.01 4.11628 to 0.01 3.88452
\PR 0.01 3.88452 to 0.03 3.88452
\PR 0.03 3.88452 to 0.03 3.72834
\PR 0.03 3.72834 to 0.05 3.72834
\PR 0.05 3.72834 to 0.05 3.30008
\PR 0.05 3.30008 to 0.07 3.30008
\PR 0.07 3.30008 to 0.07 2.75091
\PR 0.07 2.75091 to 0.09 2.75091
\PR 0.09 2.75091 to 0.09 2.32265
\PR 0.09 2.32265 to 0.11 2.32265
\PR 0.11 2.32265 to 0.11 2.06570
\PR 0.11 2.06570 to 0.13 2.06570
\PR 0.13 2.06570 to 0.13 1.55179
\PR 0.13 1.55179 to 0.15 1.55179
\PR 0.15 1.55179 to 0.15 1.07819
\PR 0.15 1.07819 to 0.17 1.07819
\PR 0.17 1.07819 to 0.17 0.71040
\PR 0.17 0.71040 to 0.19 0.71040
\PR 0.19 0.71040 to 0.19 0.53910
\PR 0.19 0.53910 to 0.21 0.53910
\PR 0.21 0.53910 to 0.21 0.31237
\PR 0.21 0.31237 to 0.23 0.31237
\PR 0.23 0.31237 to 0.23 0.21161
\PR 0.23 0.21161 to 0.25 0.21161
\PR 0.25 0.21161 to 0.25 0.12596
\PR 0.25 0.12596 to 0.27 0.12596
\PR 0.27 0.12596 to 0.27 0.08061
\PR 0.27 0.08061 to 0.29 0.08061
\PR 0.29 0.08061 to 0.29 0.02519
\PR 0.29 0.02519 to 0.31 0.02519
\PR 0.31 0.02519 to 0.31 0.01511
\PR 0.31 0.01511 to 0.33 0.01511
\PR 0.33 0.01511 to 0.33 0.01511
\PR 0.33 0.01511 to 0.35 0.01511
\PR 0.35 0.01511 to 0.35 0.01008
\PR 0.35 0.01008 to 0.37 0.01008
\PR 0.37 0.01008 to 0.37 0.01511
\PR 0.37 0.01511 to 0.39 0.01511
\PR 0.39 0.01511 to 0.39 0.00000

\put {$2\times 8^3,\xi=1.5$} at 0 5.7
\endpicture
} [lb] at -30 -24

\put { \beginpicture
\setcoordinatesystem units <120pt,24pt> point at 0 0
\setplotarea x from -0.5 to 0.5, y from -1 to 6.5
\axis left  invisible ticks in at 2 4 / short in unlabeled at 1 3 5 / /
\axis right invisible ticks in at 2 4 / short in unlabeled at 1 3 5 / /
\axis bottom shiftedto y=0 invisible ticks in numbered at -0.25 0 0.25 / /
\PR -0.5 0   to  0.5 0 \PR  0.5 0   to  0.5 6.5 \PR  0.5 6.5 to -0.5 6.5
\setlinear  \linethickness=.4pt

\PR -0.5 0.0388 to -0.49 0.03880
\PR -0.49 0.03880 to -0.49 0.07760
\PR -0.49 0.07760 to -0.47 0.07760
\PR -0.47 0.07760 to -0.47 0.05820
\PR -0.47 0.05820 to -0.45 0.05820
\PR -0.45 0.05820 to -0.45 0.04527
\PR -0.45 0.04527 to -0.43 0.04527
\PR -0.43 0.04527 to -0.43 0.09053
\PR -0.43 0.09053 to -0.41 0.09053
\PR -0.41 0.09053 to -0.41 0.11640
\PR -0.41 0.11640 to -0.39 0.11640
\PR -0.39 0.11640 to -0.39 0.14227
\PR -0.39 0.14227 to -0.37 0.14227
\PR -0.37 0.14227 to -0.37 0.15520
\PR -0.37 0.15520 to -0.35 0.15520
\PR -0.35 0.15520 to -0.35 0.20693
\PR -0.35 0.20693 to -0.33 0.20693
\PR -0.33 0.20693 to -0.33 0.27807
\PR -0.33 0.27807 to -0.31 0.27807
\PR -0.31 0.27807 to -0.31 0.41386
\PR -0.31 0.41386 to -0.29 0.41386
\PR -0.29 0.41386 to -0.29 0.54966
\PR -0.29 0.54966 to -0.27 0.54966
\PR -0.27 0.54966 to -0.27 0.52380
\PR -0.27 0.52380 to -0.25 0.52380
\PR -0.25 0.52380 to -0.25 0.75013
\PR -0.25 0.75013 to -0.23 0.75013
\PR -0.23 0.75013 to -0.23 0.85360
\PR -0.23 0.85360 to -0.21 0.85360
\PR -0.21 0.85360 to -0.21 0.93120
\PR -0.21 0.93120 to -0.19 0.93120
\PR -0.19 0.93120 to -0.19 1.37093
\PR -0.19 1.37093 to -0.17 1.37093
\PR -0.17 1.37093 to -0.17 1.56492
\PR -0.17 1.56492 to -0.15 1.56492
\PR -0.15 1.56492 to -0.15 1.96586
\PR -0.15 1.96586 to -0.13 1.96586
\PR -0.13 1.96586 to -0.13 1.87532
\PR -0.13 1.87532 to -0.11 1.87532
\PR -0.11 1.87532 to -0.11 2.25685
\PR -0.11 2.25685 to -0.09 2.25685
\PR -0.09 2.25685 to -0.09 2.26332
\PR -0.09 2.26332 to -0.07 2.26332
\PR -0.07 2.26332 to -0.07 2.75479
\PR -0.07 2.75479 to -0.05 2.75479
\PR -0.05 2.75479 to -0.05 2.74185
\PR -0.05 2.74185 to -0.03 2.74185
\PR -0.03 2.74185 to -0.03 2.85178
\PR -0.03 2.85178 to -0.01 2.85178
\PR -0.01 2.85178 to -0.01 2.82592
\PR -0.01 2.82592 to 0.01 2.82592
\PR 0.01 2.82592 to 0.01 2.90998
\PR 0.01 2.90998 to 0.03 2.90998
\PR 0.03 2.90998 to 0.03 2.38619
\PR 0.03 2.38619 to 0.05 2.38619
\PR 0.05 2.38619 to 0.05 2.44439
\PR 0.05 2.44439 to 0.07 2.44439
\PR 0.07 2.44439 to 0.07 2.18572
\PR 0.07 2.18572 to 0.09 2.18572
\PR 0.09 2.18572 to 0.09 2.08226
\PR 0.09 2.08226 to 0.11 2.08226
\PR 0.11 2.08226 to 0.11 1.73306
\PR 0.11 1.73306 to 0.13 1.73306
\PR 0.13 1.73306 to 0.13 1.58432
\PR 0.13 1.58432 to 0.15 1.58432
\PR 0.15 1.58432 to 0.15 1.35799
\PR 0.15 1.35799 to 0.17 1.35799
\PR 0.17 1.35799 to 0.17 1.11873
\PR 0.17 1.11873 to 0.19 1.11873
\PR 0.19 1.11873 to 0.19 1.02173
\PR 0.19 1.02173 to 0.21 1.02173
\PR 0.21 1.02173 to 0.21 0.72426
\PR 0.21 0.72426 to 0.23 0.72426
\PR 0.23 0.72426 to 0.23 0.58846
\PR 0.23 0.58846 to 0.25 0.58846
\PR 0.25 0.58846 to 0.25 0.56906
\PR 0.25 0.56906 to 0.27 0.56906
\PR 0.27 0.56906 to 0.27 0.32333
\PR 0.27 0.32333 to 0.29 0.32333
\PR 0.29 0.32333 to 0.29 0.29100
\PR 0.29 0.29100 to 0.31 0.29100
\PR 0.31 0.29100 to 0.31 0.28453
\PR 0.31 0.28453 to 0.33 0.28453
\PR 0.33 0.28453 to 0.33 0.19400
\PR 0.33 0.19400 to 0.35 0.19400
\PR 0.35 0.19400 to 0.35 0.12933
\PR 0.35 0.12933 to 0.37 0.12933
\PR 0.37 0.12933 to 0.37 0.13580
\PR 0.37 0.13580 to 0.39 0.13580
\PR 0.39 0.13580 to 0.39 0.06467
\PR 0.39 0.06467 to 0.41 0.06467
\PR 0.41 0.06467 to 0.41 0.05173
\PR 0.41 0.05173 to 0.43 0.05173
\PR 0.43 0.05173 to 0.43 0.05173
\PR 0.43 0.05173 to 0.45 0.05173
\PR 0.45 0.05173 to 0.45 0.03233
\PR 0.45 0.03233 to 0.47 0.03233
\PR 0.47 0.03233 to 0.47 0.02587
\PR 0.47 0.02587 to 0.49 0.02587
\PR 0.49 0.02587 to 0.49 0.05173
\PR 0.49 0.05173 to 0.50 0.05173

\put {$2\times 8^3,\xi = 2.0$} at 0 5.7
\endpicture
} [lb] at 120 -24

\put { \beginpicture
\setcoordinatesystem units <120pt,24pt> point at 0 0
\setplotarea x from -0.5 to 0.75, y from -1 to 6.5
\axis left  invisible ticks in at 2 4 / short in unlabeled at 1 3 5 / /
\axis right shiftedto x=0.5 invisible ticks in numbered  at 2 4 /
        short in unlabeled at 1 3 5 / /
\put {$\rho_{cs}$} [l] at  0.55 5.85
\axis bottom shiftedto y=0 invisible ticks in numbered at -0.25 0 0.25 / /
\PR -0.5 0   to  0.5 0 \PR  0.5 0   to  0.5 6.5 \PR  0.5 6.5 to -0.5 6.5
\setlinear  \linethickness=.4pt

\PR -0.5 0.28677 to -0.49 0.28677
\PR -0.49 0.28677 to -0.49 0.30725
\PR -0.49 0.30725 to -0.47 0.30725
\PR -0.47 0.30725 to -0.47 0.36358
\PR -0.47 0.36358 to -0.45 0.36358
\PR -0.45 0.36358 to -0.45 0.44551
\PR -0.45 0.44551 to -0.43 0.44551
\PR -0.43 0.44551 to -0.43 0.38406
\PR -0.43 0.38406 to -0.41 0.38406
\PR -0.41 0.38406 to -0.41 0.38918
\PR -0.41 0.38918 to -0.39 0.38918
\PR -0.39 0.38918 to -0.39 0.51209
\PR -0.39 0.51209 to -0.37 0.51209
\PR -0.37 0.51209 to -0.37 0.45063
\PR -0.37 0.45063 to -0.35 0.45063
\PR -0.35 0.45063 to -0.35 0.53257
\PR -0.35 0.53257 to -0.33 0.53257
\PR -0.33 0.53257 to -0.33 0.68107
\PR -0.33 0.68107 to -0.31 0.68107
\PR -0.31 0.68107 to -0.31 0.65547
\PR -0.31 0.65547 to -0.29 0.65547
\PR -0.29 0.65547 to -0.29 0.74252
\PR -0.29 0.74252 to -0.27 0.74252
\PR -0.27 0.74252 to -0.27 0.86030
\PR -0.27 0.86030 to -0.25 0.86030
\PR -0.25 0.86030 to -0.25 0.97296
\PR -0.25 0.97296 to -0.23 0.97296
\PR -0.23 0.97296 to -0.23 1.07538
\PR -0.23 1.07538 to -0.21 1.07538
\PR -0.21 1.07538 to -0.21 1.17780
\PR -0.21 1.17780 to -0.19 1.17780
\PR -0.19 1.17780 to -0.19 1.32118
\PR -0.19 1.32118 to -0.17 1.32118
\PR -0.17 1.32118 to -0.17 1.24949
\PR -0.17 1.24949 to -0.15 1.24949
\PR -0.15 1.24949 to -0.15 1.32630
\PR -0.15 1.32630 to -0.13 1.32630
\PR -0.13 1.32630 to -0.13 1.61307
\PR -0.13 1.61307 to -0.11 1.61307
\PR -0.11 1.61307 to -0.11 1.48505
\PR -0.11 1.48505 to -0.09 1.48505
\PR -0.09 1.48505 to -0.09 1.60283
\PR -0.09 1.60283 to -0.07 1.60283
\PR -0.07 1.60283 to -0.07 1.71036
\PR -0.07 1.71036 to -0.05 1.71036
\PR -0.05 1.71036 to -0.05 1.88447
\PR -0.05 1.88447 to -0.03 1.88447
\PR -0.03 1.88447 to -0.03 1.81790
\PR -0.03 1.81790 to -0.01 1.81790
\PR -0.01 1.81790 to -0.01 1.76669
\PR -0.01 1.76669 to 0.01 1.76669
\PR 0.01 1.76669 to 0.01 1.77181
\PR 0.01 1.77181 to 0.03 1.77181
\PR 0.03 1.77181 to 0.03 1.72061
\PR 0.03 1.72061 to 0.05 1.72061
\PR 0.05 1.72061 to 0.05 1.78718
\PR 0.05 1.78718 to 0.07 1.78718
\PR 0.07 1.78718 to 0.07 1.78206
\PR 0.07 1.78206 to 0.09 1.78206
\PR 0.09 1.78206 to 0.09 1.60283
\PR 0.09 1.60283 to 0.11 1.60283
\PR 0.11 1.60283 to 0.11 1.69500
\PR 0.11 1.69500 to 0.13 1.69500
\PR 0.13 1.69500 to 0.13 1.49529
\PR 0.13 1.49529 to 0.15 1.49529
\PR 0.15 1.49529 to 0.15 1.50553
\PR 0.15 1.50553 to 0.17 1.50553
\PR 0.17 1.50553 to 0.17 1.33654
\PR 0.17 1.33654 to 0.19 1.33654
\PR 0.19 1.33654 to 0.19 1.25461
\PR 0.19 1.25461 to 0.21 1.25461
\PR 0.21 1.25461 to 0.21 1.06514
\PR 0.21 1.06514 to 0.23 1.06514
\PR 0.23 1.06514 to 0.23 0.97296
\PR 0.23 0.97296 to 0.25 0.97296
\PR 0.25 0.97296 to 0.25 0.82446
\PR 0.25 0.82446 to 0.27 0.82446
\PR 0.27 0.82446 to 0.27 0.76813
\PR 0.27 0.76813 to 0.29 0.76813
\PR 0.29 0.76813 to 0.29 0.70156
\PR 0.29 0.70156 to 0.31 0.70156
\PR 0.31 0.70156 to 0.31 0.67083
\PR 0.31 0.67083 to 0.33 0.67083
\PR 0.33 0.67083 to 0.33 0.54793
\PR 0.33 0.54793 to 0.35 0.54793
\PR 0.35 0.54793 to 0.35 0.51721
\PR 0.35 0.51721 to 0.37 0.51721
\PR 0.37 0.51721 to 0.37 0.43015
\PR 0.37 0.43015 to 0.39 0.43015
\PR 0.39 0.43015 to 0.39 0.46600
\PR 0.39 0.46600 to 0.41 0.46600
\PR 0.41 0.46600 to 0.41 0.37382
\PR 0.41 0.37382 to 0.43 0.37382
\PR 0.43 0.37382 to 0.43 0.38918
\PR 0.43 0.38918 to 0.45 0.38918
\PR 0.45 0.38918 to 0.45 0.34310
\PR 0.45 0.34310 to 0.47 0.34310
\PR 0.47 0.34310 to 0.47 0.36870
\PR 0.47 0.36870 to 0.49 0.36870
\PR 0.49 0.36870 to 0.49 0.27653
\PR 0.49 0.27653 to 0.50 0.27653

\put {$2\times 8^3,\xi = 2.5$} at 0 5.7
\endpicture
} [lb] at 240 -24

\endpicture

\caption{\sl Chern-Simons number distributions on various lattices and
different anisotropies $\xi$ as indicated.
The top row gives the $T\,=\,0$ distributions at growing spatial volume.
There are only small differences at $T\!=\!T_c$ (middle row).
With increasing temperature at lattices of fixed spatial size
the distributions get significantly broader (lower row). }

\end{figure}

In Fig.\,3 we show some results for the width of the
Chern-Simons numbers, without performing any vacuum subtractions. The approach
to the limiting value of $1/12$ on large lattices is clearly visible.
It also
is obvious that on smaller lattices the fluctuations grow linearly with volume.
In fact, we find that the first two non-vanishing moments of $n_{cs}$
grow proportionally to the spatial volume and its square, respectively.
For $N_\sigma\le 20$ one obtains
\begin{eqnarray}
\langle n^2_{cs}\rangle_{\xi=1}(N_\sigma, N_\tau) & = & \cases{
(\enspace 5.9\pm 0.5)\,10^{-6}\,\, N_{\sigma}^{3}
\quad,\qquad N_\tau=N_\sigma\cr
(\enspace 8.5\pm 0.2)\,10^{-6}\,\, N_{\sigma}^{3}\quad,\qquad N_\tau=2}~~,
\nonumber \\
\langle n^4_{cs}\rangle_{\xi=1}(N_\sigma, N_\tau) & = & \cases{
(\enspace 9.6 \pm 0.6)\,10^{-11} N_{\sigma}^6 \quad, \qquad N_\tau=N_\sigma\cr
(        17.0 \pm 0.6)\,10^{-11} N_{\sigma}^6 \quad, \qquad N_\tau=2}~~.
\label{moments0}
\end{eqnarray}
Details on the statistic and the results for all the moments measured by
us on various lattices are given in Tab.\,\ref{RES}.
The small change of the distribution as function of temperature below
$T_c$ is also confirmed by our simulations on symmetric lattices of size
$8^4$ with anisotropic couplings $\xi =2$ and $4$, which corresponds to
temperatures $T/T_c \simeq 0.5$ and 1.0.

\begin{table}
\def\RL{\rule{0pt}{13pt}}
\caption{\sl The first two, non-vanishing moments of Chern-Simons number
distributions on lattices of size $N_\tau \times N_{\sigma}^3$ and
anisotropy $\xi$. Also given is the number of configurations analyzed
for each set of parameters. All gauge field configurations are
separated by 10 sweeps of overrelaxed heat bath updates.}
\label{RES}

\hskip 6pt

\begin{center} \begin{tabular}{|r|r|r|r|l|l|}
\hline
   \multicolumn{1}{|c}{$\ N_\sigma\ $\rule[-6pt]{0pt}{7pt}\RL}
 & \multicolumn{1}{|c}{$\ N_\tau\ $}
 & \multicolumn{1}{|c}{$\ \xi\ $}
 & \multicolumn{1}{|c}{\ \#\ }
 & \multicolumn{1}{|c|}{$\langle n_{cs}^2 \rangle$}
 & \multicolumn{1}{|c|}{$\langle n_{cs}^4 \rangle \cdot $} \\
\hline
 2\ \RL &  2\ & \ 1.0\ & \ 8000\ & \ 0.000033(2)\ & \ 0.0000000046(04)\ \\
 2\ \   &  2\ & \ 2.0\ & \ 8000\ & \ 0.000104(5)\ & \ 0.0000000421(30)\ \\
\hline
 4\ \RL &  4\ & \ 1.0\ & \ 8400\ & \ 0.000334(10)\ & \ 0.00000035(2)\ \\
 4\ \   &  2\ & \ 1.0\ & \ 9840\ & \ 0.000443(15)\ & \ 0.00000062(3)\ \\
 4\ \   &  2\ & \ 1.2\ & \ 8000\ & \ 0.000622(20)\ & \ 0.00000122(6)\ \\
 4\ \   &  2\ & \ 1.5\ & \ 1200\ & \ 0.000966(50)\ & \ 0.0000029(4)\ \\
 4\ \   &  2\ & \ 1.6\ & \ 8000\ & \ 0.001016(30)\ & \ 0.0000034(2)\ \\
 4\ \   &  2\ & \ 1.7\ & \ 8000\ & \ 0.001132(30)\ & \ 0.0000041(2)\ \\
 4\ \   &  2\ & \ 1.8\ & \ 8000\ & \ 0.001294(30)\ & \ 0.0000056(3) \ \\
 4\ \   &  2\ & \ 2.0\ & \ 1000\ & \ 0.00173(20)\  & \ 0.00000875(150)\ \\
 4\ \   &  2\ & \ 2.2\ & \ 6628\ & \ 0.002021(50)\ & \ 0.0000136(10)\ \\
 4\ \   &  2\ & \ 2.4\ & \ 6848\ & \ 0.002399(60)\ & \ 0.0000191(15)\ \\
 4\ \   &  2\ & \ 3.0\ & \ 1000\ & \ 0.00480(14)\  & \ 0.0000867(80)\ \\
\hline
 6\ \RL &  6\ & \ 1.0\ & \ 1560\ & \ 0.00123(7)\   & \ 0.0000049(005)\ \\
 6\ \   &  2\ & \ 1.0\ & \ 7680\ & \ 0.00175(6)\   & \ 0.0000094(005)\ \\
 6\ \   &  2\ & \ 1.2\ & \ 4036\ & \ 0.00242(7)\   & \ 0.0000189(12)\ \\
 6\ \   &  2\ & \ 1.5\ & \ 2058\ & \ 0.0038(2)\    & \ 0.0000447(35)\ \\
 6\ \   &  2\ & \ 1.8\ & \ 4160\ & \ 0.0054(2)\    & \ 0.0000946(90)\ \\
 6\ \   &  2\ & \ 2.0\ & \ 7760\ & \ 0.0070(3)\    & \ 0.000155(10)\ \\
 6\ \   &  2\ & \ 2.2\ & \ 6914\ & \ 0.0101(4)\    & \ 0.00370(40)\ \\
 6\ \   &  2\ & \ 2.5\ & \ 5472\ & \ 0.0128(4)\    & \ 0.00504(30)\ \\
\hline
 8\ \RL &  8\ & \ 1.0\ & \ 2400\ & \ 0.00284(13)\  & \ 0.000024(2)\  \\
 8\ \   &  8\ & \ 2.0\ & \ 7840\ & \ 0.00468(15)\  & \ 0.000065(4)\  \\
 8\ \   &  8\ & \ 4.0\ & \ 2952\ & \ 0.00320(17)\  & \ 0.000032(4)\  \\
 8\ \   &  2\ & \ 1.0\ & \ 9120\ & \ 0.00426(13)\  & \ 0.000054(4)\  \\
 8\ \   &  2\ & \ 1.3\ & \ 2004\ & \ 0.00731(60)\  & \ 0.000169(20)\ \\
 8\ \   &  2\ & \ 1.5\ & \ 9924\ & \ 0.00976(40)\  & \ 0.000293(20)\ \\
 8\ \   &  2\ & \ 1.8\ & \ 3444\ & \ 0.01539(110)\ & \ 0.000729(70)\ \\
 8\ \   &  2\ & \ 2.0\ & \ 7732\ & \ 0.02241(100)\ & \ 0.00162(20)\  \\
 8\ \   &  2\ & \ 2.5\ & \ 18564\ & \ 0.04621(200)\ & \ 0.00545(30)\  \\
\hline
10\ \RL &  2\ & \ 1.0\ & \ 2720\ & \ 0.00864(50)\  & \ 0.00023(2)\ \\
\hline
12\ \RL &  2\ & \ 1.0\ & \ 1860\ & \ 0.0151(11)\  & \ 0.00065(7)\ \\
12\ \   &  2\ & \ 2.0\ & \ 1830\ & \ 0.0721(55)\  & \ 0.0103(7)\  \\
\hline
20\ \RL &  2\ & \ 1.0\ & \ 1352\ & \ 0.0618(40)\ & \ 0.0083(5)\ \\
20\ \   &  2\ & \ 2.0\ & \ 1156\ & \ 0.0830(50)\ & \ 0.0124(5)\ \\
\hline
\end{tabular} \end{center}

\end{table}

In order to obtain information at even higher temperature in the
symmetric phase, we used anisotropic lattices with temporal extent
$N_{\tau} = 2$. Some results for anisotropy $\xi\!=\!1.5$, 2 and 2.5,
corresponding to $T/T_c=1.5,~2$ and 2.5, are shown in the last row of Fig.\,2.
The rapid broadening of the distributions with increasing $\xi$ is clearly
visible.
The resulting width of the distribution for anisotropy $\xi=2$,
corresponding to $T\simeq 2T_c$, is
also given in Fig.\,3 together with results for
$T\simeq 0$ and $T_c$. For $\xi=2$ we reach the limit of a flat
distribution,
resulting in $\langle n_{cs}^2\rangle_\xi = 1/12$, for $N_\sigma \simeq
12$, ie. $LT =12$. Our numerical simulations thus have to stay in the
regime with $LT \le 10$. As discussed in the previous section we should
compare in this parameter regime our numerical data with perturbation
theory in a finite volume.

\begin{figure}
\def\PR{\putrule from }
\beginpicture

\setcoordinatesystem units <1pt,1pt>
\setcoordinatesystem point at 0 0
\setplotarea x from -10 to 360, y from 0 to 40

\put { \beginpicture

\setcoordinatesystem units <35pt,21pt>
\setcoordinatesystem point at 0 0
\setplotarea x from 1 to 10, y from -11 to -1.5

\axis left ticks in
      withvalues {$\displaystyle \frac{1}{12}$} 0.01    0.001   0.0001 /
              at -2.4849                       -4.6052 -6.9078 -9.2103 / /
\put {$\langle n_{cs}^2\rangle$} [r] at 0.72 -1.4

\axis bottom ticks in
  withvalues {$2^3$} {$4^3$} {$6^3$} {$8^3$} {$12^3$} {$20^3$} /
          at 2.0794  4.1589  5.3752  6.2383  7.4547   8.9872 / /
\put {$N_\sigma^3$} at 10.2 -11.6

\setlinear  \linethickness=.4pt

\put {$\circ$}      at 8.5 -9.0
\put {$+$}     at 8.5 -8.3
\put {$\times$}          at 8.5 -7.6
\put {$T=0$}    [l] at 9   -9.0
\put {$T=T_c$}  [l] at 9   -8.3
\put {$T=2T_c$} [l] at 9   -7.6

\setdashpattern <4pt,4pt>
\plot 2.0794  -10.3137 4.1589   -8.0043 5.3752   -6.6926 6.2383   -5.8657 /
\multiput {$\circ$} at   2.0794  -10.3137
 4.1589   -8.0043 5.3752   -6.6926 6.2383   -5.8657 /
\setdashpattern <6pt,3pt,2pt,3pt>
\plot 2.0794  -10.3137 4.1589   -7.7221 5.3752   -6.3503 6.2383   -5.4579
        6.9078   -4.7519 7.4547   -4.1943 8.9872   -2.7832 /
  \multiput {$+$} at  2.0794  -10.3137 4.1589   -7.7221 5.3752   -6.3503
      6.2383   -5.4579 6.9078   -4.7519 7.4547   -4.1943 8.9872   -2.7832 /
\setdashpattern <8pt,4pt,2pt,4pt>
\plot 2.0794   -9.1717 5.3752   -4.9602 6.2383   -3.8035
        7.4547   -2.6300 8.9872   -2.4890 /
\multiput {$\times$}  at  2.0794   -9.1717 5.3752   -4.9602
        6.2383   -3.8035 7.4547   -2.6300 8.9872   -2.4890 /
\setsolid
\PR 1 -2.4849 to 10 -2.4849
\endpicture
} [lb] at 0 0

\endpicture

\caption{\sl Width of the Chern-Simons number distributions
as a function of spatial volume. }

\end{figure}
\begin{figure}
\beginpicture

\setcoordinatesystem units <1pt,1pt>
\setcoordinatesystem point at 0 0
\setplotarea x from -10 to 360, y from 0 to 40

\put { \beginpicture

\setcoordinatesystem units <160pt,30pt>
\setcoordinatesystem point at 0 0
\setplotarea x from -0.3 to 1.2, y from -8 to -2

\axis left ticks in
      withvalues  0.001   0.002   0.005   0.01    0.02    0.05 /
              at -6.9078 -6.2146 -5.2983 -4.6052 -3.9120 -2.9957 / /
\put {$\xi$} at 1.14 -8.4
\axis bottom ticks in withvalues  1  1.5 2 2.5 /
                  at  0  .4055 .6931 0.9163 / /
\put {$\langle n_{cs}^2\rangle$} [r] at -0.34 -2.3

\multiput {$\circ$} at   0      -5.4579 0.2623 -4.9184 0.4055 -4.6298
            0.5878 -4.1740 0.6931 -3.8035 0.9163 -3.0746 /

  \multiput {$\times$} at  0      -6.557 0.2623 -5.591 0.4055 -5.171
                           0.5878 -4.514 0.6931 -4.039 0.9163 -3.196 /
\errbar  0      -6.557  0.18
\errbar  0.2623 -5.591  0.16
\errbar  0.4055 -5.171  0.1
\setdots
\accountingoff
\setdashpattern <4pt,4pt>
\plot 0      -7.7533 0.2231 -6.9926 0.4055 -6.3880 0.5596 -5.8859
      0.6931 -5.4564 0.8109 -5.0810 0.9163 -4.7476 /
\multiput {\fiverm .} at  0.025 -7.8949 0.05 -7.8199 0.075 -7.7449
  0.1 -7.6699 0.125 -7.5949 0.15 -7.5199 0.175 -7.4449 0.2 -7.3699
  0.225 -7.2949 0.25 -7.2199 0.275 -7.1449 0.3 -7.0699 0.325 -6.9949
  0.35 -6.9199 0.375 -6.8449 0.4 -6.7699 0.425 -6.6949 0.45 -6.6199
  0.475 -6.5449 0.5 -6.4699 0.525 -6.3949 0.55 -6.3199 0.575 -6.2449
  0.6 -6.1699 0.625 -6.0949 0.65 -6.0199 0.675 -5.9449 0.7 -5.8699
  0.725 -5.7949 0.75 -5.7199 0.775 -5.6449 0.8 -5.5699 0.825 -5.4949
  0.85 -5.4199 0.875 -5.3449 0.9 -5.2699 0.925 -5.1949 0.95 -5.1199
  0.975 -5.0449 /
\accountingon

\endpicture
} [lb] at 40 0

\endpicture

\caption{\sl The temperature dependence of the width on $8^3\!\times\!2$
lattices with anisotropies $\xi$ ranging from $1.0$ up to $2.5$ ($\circ$).
Subtraction of the appropiate zero temperature contribution gives the values
($\times$).  Errors are plotted only if they are bigger then the symbol.
The dashed and dotted curves give the perturbative results in the continuum
and on the lattice, respectively. }

\end{figure}

A systematic analysis of the temperature dependence of
the width of the distributions has been performed on $2 \times 8^3$ lattices
with anisotropies varying between $\xi = 1.0$ and $\xi = 2.5\,$.
With increasing temperature the distributions become rapidly broader.
Results for the thermal part of the width, calculated according to the
prescription given in Eq.(\ref{ncslat}), are shown in
Fig.\,4 together with the perturbative result.
The width clearly rises rapidly with $\xi$.
The double logarithmic plot nicely shows a power law
behaviour, $\langle n_{cs}^2 \rangle \sim \xi^{\alpha}$. The best fit
gives a power of $\alpha = 3.7(1)$, which should be compared to the
temperature dependence,
$\langle n_{cs}^2 \rangle \sim T^3$, expected from continuum
perturbation theory. Here one has to take into account, that we have
ignored quantum corrections in the lattice anisotropy couplings
$\gamma_{g,h}$, which modify the tree level relation,
$\gamma_{g,h} = \xi \sim T$,
between these couplings and the temperature.
For instance, assuming for the $O(g^2)$ correction to the anisotropic
couplings the form,
$\gamma_{g,h} = \xi (1 + c\,g^2\,(\xi\!-\!1))$, which is valid for $\xi$
close to 1,
will reduce the power to $3.2(1)$ for $cg^2=0.1$ and $2.9(1)$ for $cg^2=0.2$.
We thus consider our result for the $\xi$ dependence of the width as
rather satisfactory.

\setlength{\parskip}{1ex}

Besides this agreement in the functional form of the temperature
dependence we find, however, that the thermal width
of the Chern-Simons number distributions generally is about a factor
3.0 larger than the perturbative value. For
$\xi \ge 1.8 $ we also find a statistically significant probability for
half integer
Chern-Simons numbers. This allows us to attempt an estimate of the temperature
dependence of the tunneling rate between different Chern-Simons vacuums.
Assuming that the number of tunneling events is proportional to the number of
configurations with Chern-Simons numbers close to half-integer values, i.e.
$n_{cs} \in [n+1/2-\varepsilon, n+1/2+\varepsilon]\,,n\epsilon Z$,
we can compare the ratios of
tunneling rates at different temperatures (Tab.\,\ref{TUN}).

\begin{table}
\def\RL{\rule{0pt}{13pt}}
\caption{\sl The first table gives the fraction Chern-Simons numbers,
$F(\varepsilon,\xi)$, calculated on $2\times 8^3$ lattices with anisotropy
$\xi$ which differ from half-integer values by less than $\varepsilon$.
In the second table we compare some ratios, $F(\varepsilon,\xi_1)/
F(\varepsilon,\xi_2)$, with the semi-classical estimate, Eq.\,(4.3),
for the tunneling rate in the two limiting cases $m_W  = m_W (T\!=\!0)$ (a)
and $m_W\sim T$ (b), respectively.}
\label{TUN}

\hskip 6pt

\begin{center} \begin{tabular}{|r|r|r|}
  \hline
     \multicolumn{1}{|c}{$\xi $\rule[-6pt]{0pt}{7pt}\RL}
   & \multicolumn{1}{|c}{$F(0.1,\xi)$}
   & \multicolumn{1}{|c|}{$F(0.05,\xi)$} \\
  \hline
{\RL}1.8 &  0.0035(13) & 0.0013(07) \\
     2.0 &  0.0109(23) & 0.0044(11) \\
     2.5 &  0.0710(45) & 0.0327(28) \\
  \hline
\end{tabular} \end{center}

\begin{center} \begin{tabular}{|r|r|r|r|r|}
  \hline
     \multicolumn{1}{|c}{$\xi_1/\xi_2$\rule[-6pt]{0pt}{7pt}\RL}
   & \multicolumn{1}{|c}{$\varepsilon=0.1$}
   & \multicolumn{1}{|c}{$\varepsilon=0.05$}
   & \multicolumn{1}{|c}{$ (a)$}
   & \multicolumn{1}{|c|}{$ (b)$} \\
  \hline
{\RL}2.0/1.8  & $3.1 \pm 1.3$  & $3.4 \pm 2.0$  &  3.9  & 1.5 \\
     2.5/2.0  & $6.5 \pm 1.4$  & $7.4 \pm 2.0$  & 10.0  & 2.4 \\
  \hline
\end{tabular} \end{center}

\end{table}

We define $F(\varepsilon, \xi)$ as the fraction of configurations on
lattices with anisotropy $\xi$ (temperature $a_\sigma T = \xi/N_\tau$),
for which the absolute value of $n_{cs}$ differs by less then $\varepsilon$
from 1/2. At zero temperature we do not find any configurations with
Chern-Simons numbers close to 1/2. We thus can directly use the results
obtained on the $N_\tau =2$ lattices to determine $F(\varepsilon, \xi)$. The
data are given in Tab.\,\ref{TUN}. From this we can eliminate the unknown
proportionality constant, which relates $F$ to the tunneling rate $\Gamma$.

As long as the temperature is much smaller than the W-boson mass
the latter determines the scale for finite energy sphaleron solutions
which enter the semi-classical estimates of the
tunneling rate \cite{ArnMcLe}. One finds
exponentially small tunneling rates per unit time and volume,
\begin{equation}
\Gamma/tV = 0.007 (\alpha_wT)^4 \left(\frac{3M_W}{T\alpha_w}\right)^7
e^{-3M_W/T\alpha_w}\ .
\label{trates}
\end{equation}
At high temperatures the relevant energy scales are $\sim T$, which should lead
to a replacement of $m_W$ by a term $\sim T$
in the above estimate. One thus would expect that in the high temperature limit
the tunneling rates become proportional to $T^4$ \cite{ArnMcLe,Shap}.
In order to compare the temperature dependence of the tunneling rates
found in our numerical simulation
with the above semi-classical relation we determine the temperature and the
coupling $\alpha_w$ from our simulation parameters as
$T=\xi/N_\tau a_\sigma$ and $\alpha_w=1/\beta\pi$, respectively.
For the mass scale, $m_W$
we consider the two extreme cases $m_W \sim T$ and
$m_W= m_W(T\!=\!0)$, where
$a_\sigma M_W(T\!=\!0) = 0.2$ is taken in accordance with the
Monte Carlo simulation of Ref.\cite{Bunk}.
As can be seen from Tab.\,\ref{TUN} in the temperature regime studied
by us the data seem to favour a
mass scale, which still is only weakly temperature dependent.
This is, in fact, consistent
with the findings of Ref.\cite{Bunk}, where little temperature dependence
has been observed for the W-boson mass across the phase transition, while the
Higgs boson mass dropped significantly close to $T_H$.
Due to the low statistics the errors on our numerical results for the tunneling
rates are still quite large. However, it is
reasuring that the results do not seem to depend much on the value chosen
for $\varepsilon$.

Finally we want to test, to what extent the static approximation used
in analytical approaches is supported by our 4-dimensional simulations.
Static configurations should display strong correlations between the
Chern-Simons numbers calculated on neighboring timeslices. This is easily
visualized,
if the Chern-Simons numbers calculated on the two timeslices of our
$N_\tau = 2$ lattices are plotted against each other (Fig.\,5).
At low temperatures both numbers are uncorrelated, resulting in a spherical
distribution in the scatter plot. With increasing temperature, however, the
two measurements get more and more correlated.

The same behaviour appears when comparing the results
from simulations on different lattices which yield a
similar width but correspond to
different temperatures, e.\,g. $6^3\times 2,\ \xi\!=\!2$ and
$8^3\times 2,\ \xi\!=\!1$ (Fig.\,5).
The strength of this correlations can be measured by the covariance matrix
$cov = (\langle xy \rangle - \langle x \rangle \langle y \rangle )
/ \sqrt{(\langle x^2 \rangle \!-\! \langle x \rangle^2)
   (\langle y^2 \rangle \!-\! \langle y \rangle^2)}$,
with $x,y$ denoting
the two Chern-Simons numbers.
Due to the periodic structure of the Chern-Simons term
some care has to be taken when projecting two Chern-Simons numbers
simultaneously to the restricted interval $[-1/2,1/2]$, as artificial
correlations can buildt up.
On the other hand existing correlations may also be destroyed
because pairs of the form
$(0.5+\varepsilon, 0.5-\varepsilon)$ become separated after projection.
Therefore the numbers were shifted in pairs, minimizing
the distance from $(0,0)\,$.
Taking this definition there is a clear indication for a growing
correlation with increasing temperature (Fig.\,6).

\begin{figure}

\epsfbox[85 70 200 400]{scatter1.ps}

\epsfbox[85 70 200 280]{scatter2.ps}

\caption{\sl The scattering of Chern-Simons numbers the two timeslices is
given for fixed lattice and increasing temperature (top). Below the numbers
are displayed for simulations with similar width but different temperature. }

\end{figure}
\begin{figure}
\beginpicture

\setcoordinatesystem units <1pt,1pt>
\setcoordinatesystem point at 0 0
\setplotarea x from -10 to 360, y from 0 to 40

\put { \beginpicture

\setcoordinatesystem units <100pt,160pt>
\setcoordinatesystem point at 0 0
\setplotarea x from 0.5 to 2.7, y from 0 to 1.14

\axis left ticks in numbered at 0.0 0.2 0.4 0.6 0.8 1.0 / /
\put {cov} [r] at 0.4 1.11
\axis bottom ticks in numbered at  1  1.5 2 2.5 / /
\put {$\xi$} at 2.69 -0.07

\setlinear  \linethickness=.4pt

\accountingoff
\multiput {$\circ$} at    1.0  0.2752 1.3  0.4530 1.5  0.5631
                          1.8  0.6887 2.0  0.7657 2.5  0.8685 /
\setsolid
\plot  1.0  0.2752 1.3  0.4530 1.5  0.5631
       1.8  0.6887 2.0  0.7657 2.5  0.8685 /
\errbar  1.0  0.2752 0.015
\errbar  1.3  0.4530 0.033
\errbar  2.0  0.7657 0.013
\setdashes\setquadratic
\plot 0.7  0.2145 0.8  0.2560 0.9  0.2972 1.0  0.3376 1.1  0.3768
      1.2  0.4145 1.3  0.4505 1.4  0.4846 1.5  0.5168 1.6  0.5470
      1.7  0.5753 1.8  0.6016 1.9  0.6261 2.0  0.6489 2.1  0.6701
      2.2  0.6897 2.3  0.7079 2.4  0.7248 2.5  0.7404 /
\plot 0.7  0.1228 0.8  0.1390 0.9  0.1550 1.0  0.1710 1.1  0.1870
      1.2  0.2033 1.3  0.2196 1.4  0.2361 1.5  0.2526 1.6  0.2692
      1.7  0.2857 1.8  0.3022 1.9  0.3185 2.0  0.3347 2.1  0.3506
      2.2  0.3664 2.3  0.3819 2.4  0.3970 2.5  0.4120 /

\accountingon

\endpicture
} [lb] at 80 0

\endpicture

\caption{\sl The correlation between the two Chern-Simons numbers on the
$2\times 8^3$ lattices versus anisotropy $\xi$.
 As a measure for it we used the covariance.
 The curves give the perturbative values on the lattices with
 $N_\tau=2$ (top) and $N_\tau=N_\sigma$ (bottom), respectively.
 Error bars are only drawn if they are bigger than the symbol. }

\end{figure}

\section{Conclusions}

\setlength{\parskip}{2ex}

We have studied the temperature dependence of Chern-Simons number
distributions on Euclidean lattices.
In our parameter range we were able to produce statistically significant
distributions of Chern-Simons numbers, which clearly showed the expected
broadening of the distributions with increasing temperature.

A comparison with perturbative calculations
on the lattice as well as in the continuum shows that
the width of these distributions typically is about a factor three larger
than expected from perturbation theory. For temperatures $T \ge 1.8T_H$ we
find statistically significant fractions of configurations with Chern-Simons
numbers close to $\pm 1/2$. These configurations have been related to the
number of
tunnelings between topologically distinct vacuums. We find that the
corresponding
tunneling rates are still controlled by an energy scale consistent with that
of the
zero temperature W-boson mass. The rates do, however, start
growing rapidly between $T=T_H$ and $T=2.5T_H$, while they show little
temperature
dependence below $T_H$ as can be deduced from the small changes in the first
two, non-vanishing moments of the Chern-Simons number distributions.

Our present analysis is limited to a temperature and
volume range given by the constraint $LT < 10$, which essentially is dictated
by the occurrence of large contributions from vacuum fluctuations. If we
want to reach even higher temperatures on larger lattices, we have to perform
simulations at smaller values of the gauge coupling. With our present
algorithms this
should be feasible and it should then be possible to perform a systematic
study of the temperature dependence of the tunneling rates over a wide
temperature regime. It would certainly be interesting to check at which
temperatures one reaches a regime where
the asymptotically expected scaling of the transition rate with the fourth
power of the temperature is valid.

{\bf Acknowledgements:}
The computations have been performed on the \mbox{NEC SX-3} of the University
of K\"oln. We thank in particular J.\,Boll for his support.
We also thank Sourendu Gupta for helpful discussions.

\end{document}